\pdfoutput=1

\documentclass[twocolumn]{aastex631}
\usepackage{epstopdf}
\usepackage{xspace}
\usepackage{xcolor}
\usepackage{graphicx}
\usepackage{amssymb}
\usepackage{amsmath}
\usepackage{natbib}
\usepackage{float}
\usepackage{hyperref}
\shorttitle{Multiwavelength Study of DC\,314.8--5.1}
\shortauthors{Kosmaczewski et al.}

\begin{document}

\title{Multiwavelength Study of Dark Globule DC\,314.8--5.1:\\ Point Source Identification and Diffuse Emission Characterization}

\correspondingauthor{E.~Kosmaczewski}
\email{emily@oa.uj.edu.pl}

\author{E.~Kosmaczewski}
\affiliation{Astronomical Observatory of the Jagiellonian University, ul. Orla 171, 30-244 Krak\'ow, Poland}

\author{{\L}.~Stawarz}
\affiliation{Astronomical Observatory of the Jagiellonian University, ul. Orla 171, 30-244 Krak\'ow, Poland}

\author{C.~C.~Cheung}
\affiliation{Naval Research Laboratory, Space Science Division, Washington, DC 20375, USA}

\author{A. Bamba}
\affiliation{Department of Physics, Graduate School of Science, The University of Tokyo, 7-3-1 Hongo, Bunkyo-ku, Tokyo 113-0033, Japan}
\affiliation{Research Center for the Early Universe, School of Science, The University of Tokyo, 7-3-1 Hongo, Bunkyo-ku, Tokyo 113-0033, Japan }
\affiliation{Trans-Scale Quantum Science Institute, The University of Tokyo, Tokyo 113-0033, Japan}

\author{A.~Karska}
\affiliation{Max-Planck-Institut f\"ur Radioastronomie, Auf dem H\"ugel 69, 53121, Bonn, Germany}
\affiliation{Institute of Astronomy, Faculty of Physics, Astronomy and Informatics, Nicolaus Copernicus University, ul. Grudziadzka 5, 87-100 Toru\'n, Poland}

\author{W.~R.~M.~Rocha}
\affiliation{Laboratory for Astrophysics, Leiden Observatory, Leiden University, P.O. Box 9513, NL 2300 RA Leiden, The Netherlands}

\begin{abstract}
We present an analysis of multi-wavelength observations of the dark globule DC\,314.8--5.1, using data from the Gaia optical, 2MASS near-infrared, and WISE mid-infrared surveys, dedicated imaging with the Spitzer Space Telescope, and X-ray data obtained with the Swift-XRT Telescope (XRT). The main goal was to identify possible pre-main sequence stars (PMSs) and young stellar objects (YSOs) associated with the globule. For this, we studied the infrared colors of all point sources within the boundaries of the cloud. After removing sources with non-stellar spectra, we investigated the Gaia parallaxes for the YSO candidates, and found that none are physically related to DC\,314.8--5.1. In addition, we searched for X-ray emission from pre-main sequence stars with Swift-XRT, and found no 0.5--10\,keV emission down to a luminosity level $\lesssim 10^{31}$erg\,s$^{-1}$,  typical of a PMS with mass\,$\ge 2 M_\odot$. Our detailed inspection therefore supports a very young, ``pre-stellar core'' evolutionary stage for the cloud. Based on archival Planck and IRAS data, we moreover identify the presence of hot dust, with temperatures $\gtrsim 100$\,K, in addition to the dominant dust component at 14\,K, originating with the associated reflection nebula.
\end{abstract} 

\section{Introduction} 
\label{sec:intro}

The physical state of molecular clouds at a given evolutionary stage is strongly dependent on the development of star formation within such systems \citep[for a review, see e.g.;][]{Heyer15, Klessen16, Jorgensen20}. The interactions of young stellar objects (YSOs) with their host clouds are substantial at the early stages of stellar formation. Stars form when the dense cores of these clouds collapse, with the infall of material resulting in the gravitational potential energy heating the material and increasing its density up to $\sim 10^8-10^9$\,cm$^{-3}$ \citep{Jorgensen20}. The main effects of stellar formation are the processing of the dust within the cloud, the disruption of the cloud structure, and heating of the cloud material \citep{Strom75}. These processes continue as the system is altered and disrupted by the evolving young star.

Consequently, there is much interest in studying clouds prior to the onset of star formation, in particular pre-stellar cores \citep[see][for a review]{Bergin07}. Much study has been done on the already known pre-stellar cores, TMC-1 and L134N, however much of this was restricted to the sub-mm and radio end of the electromagnetic spectrum \citep{Bergin07}. \citet{Kirk05} did a survey for pre-stellar cores, detecting 29 cores with sub-mm observations, and established some basic characteristics expected of such pre-stellar cores, such as an average temperature of 10\,K, volume densities of bright cores of $10^5-10^6$\,cm$^{-3}$ and intermediate cores of $10^4-10^5$\,cm$^{-3}$, and additionally constrained radial density profiles and lifetimes of such cores. The filamentary structures of molecular clouds down to the internal structures of dense cores was further studied by \citet{Andre14}.

It is with this context in mind that we examined the pre-stellar nature of the dark globule DC\,314.8--5.1. Originally classified as a compact dark globule \citep{Hartley86}, it has a serendipitous association with a field star which illuminates reflection from the cloud. \citet{Whittet07} concluded, from Infrared Astronomical Satellite \citep[IRAS;][]{IRAS} and Two Micron All Sky Survey \citep[2MASS;][]{2MASS} data, that the cloud is at the onset of low-mass star formation and further discussed the basic properties of the system. \citeauthor{Whittet07} performed a 2MASS survey of an elliptical region with radii $7^\prime \times 5^\prime$ covering the extent of the cloud to identify potential YSOs, and found only two candidates out of the sample of 387 sources. One was excluded as an old star with significant dust reddening and the other remained a viable YSO candidate, hereafter referred to as ``C1.'' 

In \citet{Kosmaczewski22}, we showed a presence of divergent conditions for DC\,314.8-5.1 as compared to molecular clouds with ongoing star formation. In particular, our study of the Spitzer Space Telescope InfraRed Spectrograph \citep[IRS;]{Houck04} mid-infrared spectra revealed a surprisingly large cation-to-neutral PAH ratio, which could be explained by lower-energy cosmic rays (CRs) ionizing the cloud's interior, in addition to photo-ionization by the star. However, to confirm this, one must perform a deeper search, to rule out pre/young-stellar objects.

In this paper, we expand on YSO identification methods for this system, utilizing data from dedicated observations with Spitzer and the Neil Gehrels Swift Observatory \citep[Swift;][]{Burrows00}, as well as from archival Wide-field Infrared Survey Explorer \citep[WISE;][]{Wright10}, 2MASS, and Gaia \citep{Gaia21} surveys. The main goals of the multi-wavelength data analysis presented here are: (i) to identify YSO candidates utilizing infrared and optical imaging, (ii) to test for the presence of pre-main sequence stars (PMSs) that exhibit no optical/infrared counterparts, and (iii) to characterize the diffuse emission seen at microwave and infrared frequencies.

\begin{figure}[th!]
\centering
\includegraphics[width=0.49\textwidth]{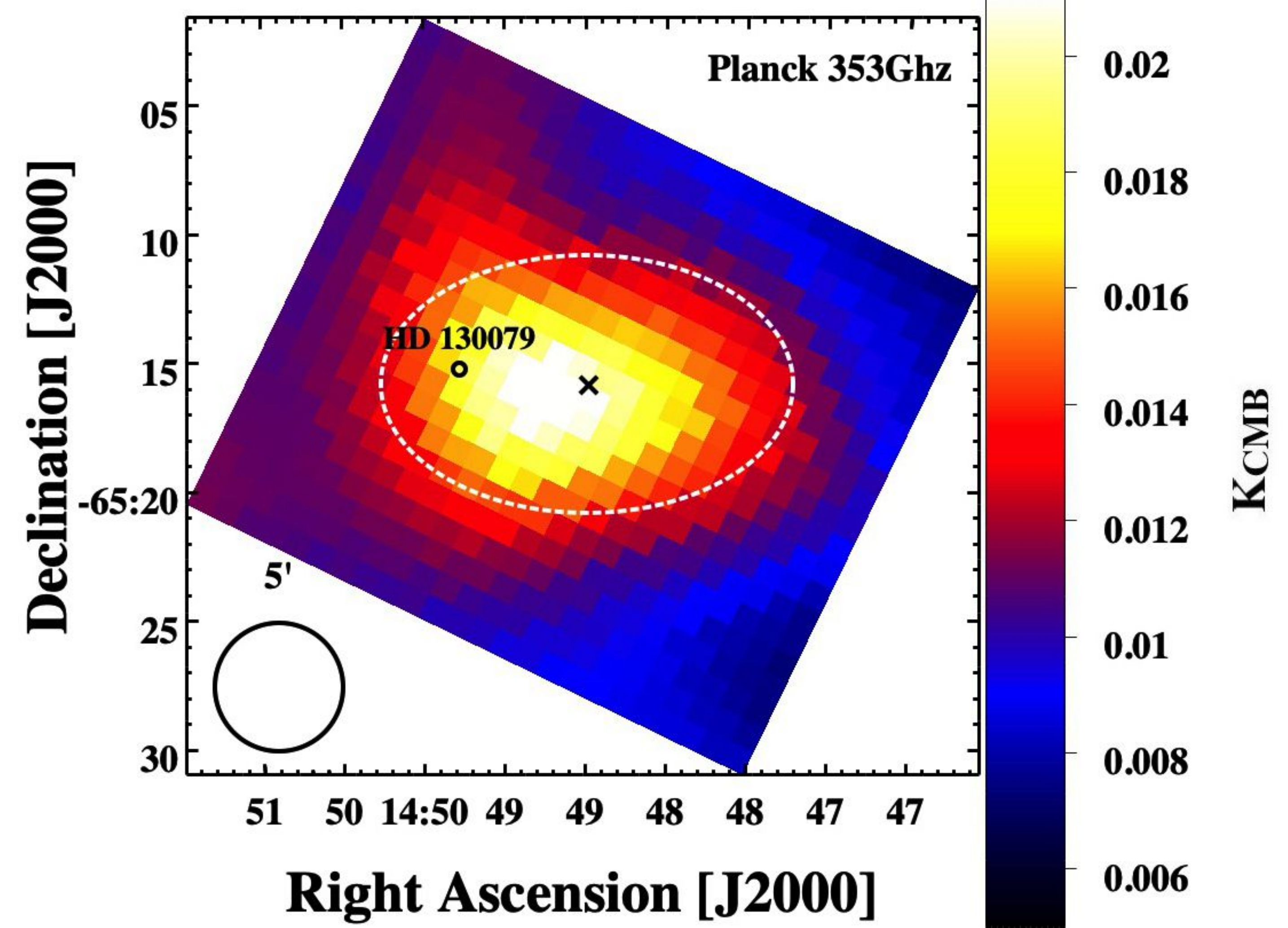}
\includegraphics[width=0.49\textwidth]{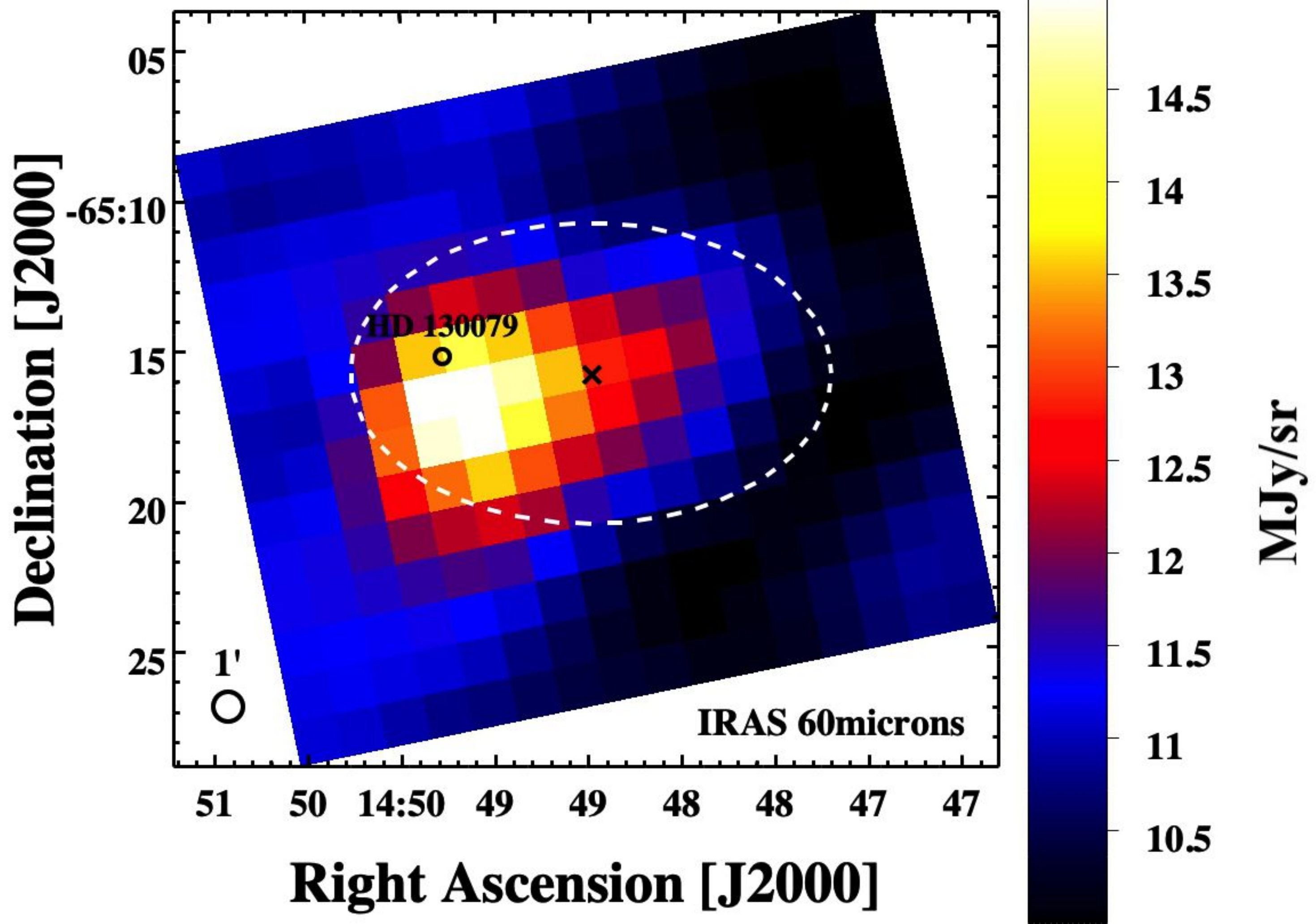}
\caption{DC\,314.8--5.1 as seen by  Planck  at 353\,GHz (top panel) and IRAS at 60\,$\mu$m (bottom panel). The white dashed ellipse (with radii of $\sim 7^{\prime} \times 5^{\prime}$) denotes the globule,} with the central position marked by a black ``x''. The star HD\,130079 is marked with a black open circle to the east of the cloud center.
\label{fig:low-res}
\end{figure}

\begin{figure*}[th!]
\centering
\includegraphics[width=0.49\textwidth]{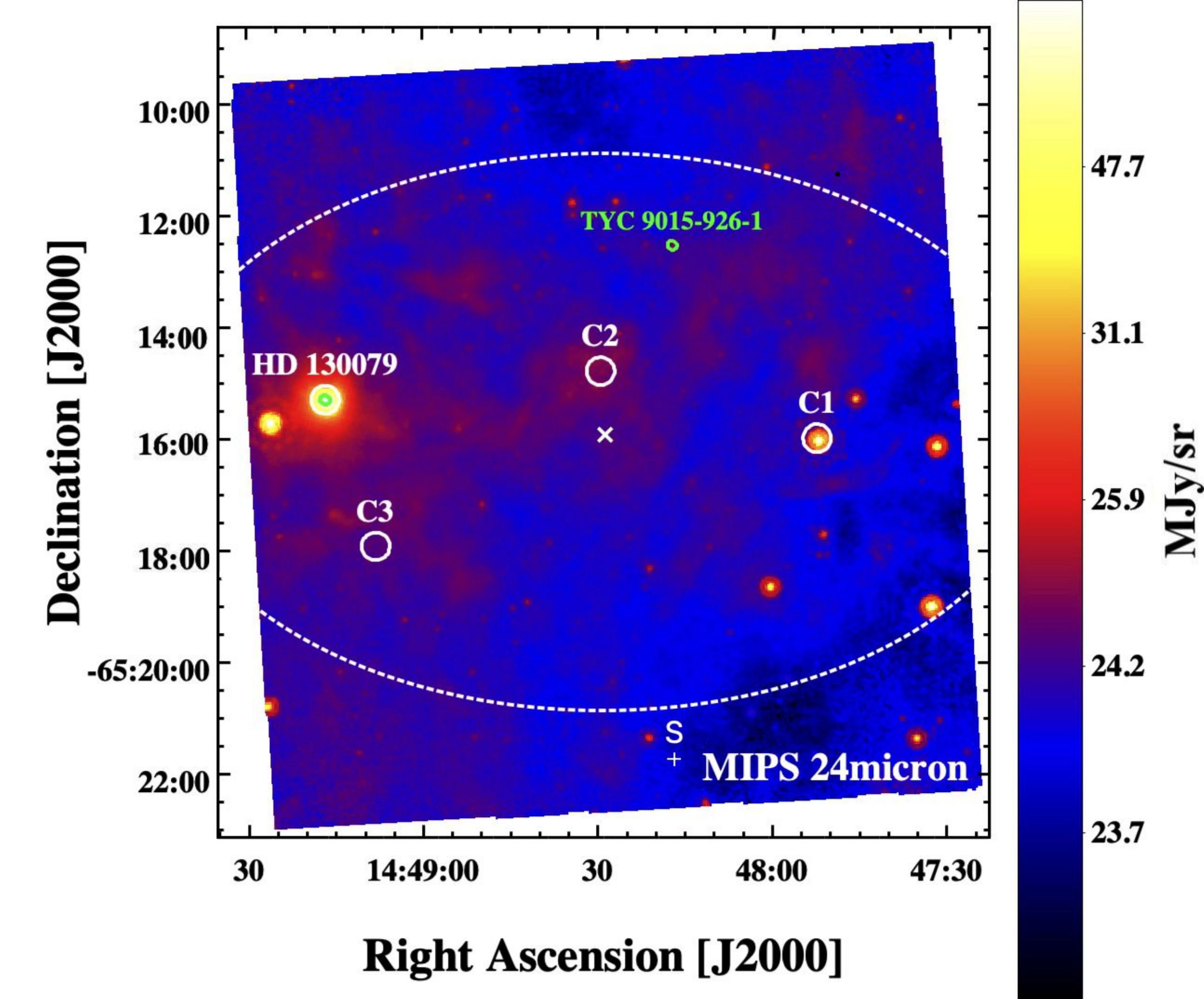}
\includegraphics[width=0.49\textwidth]{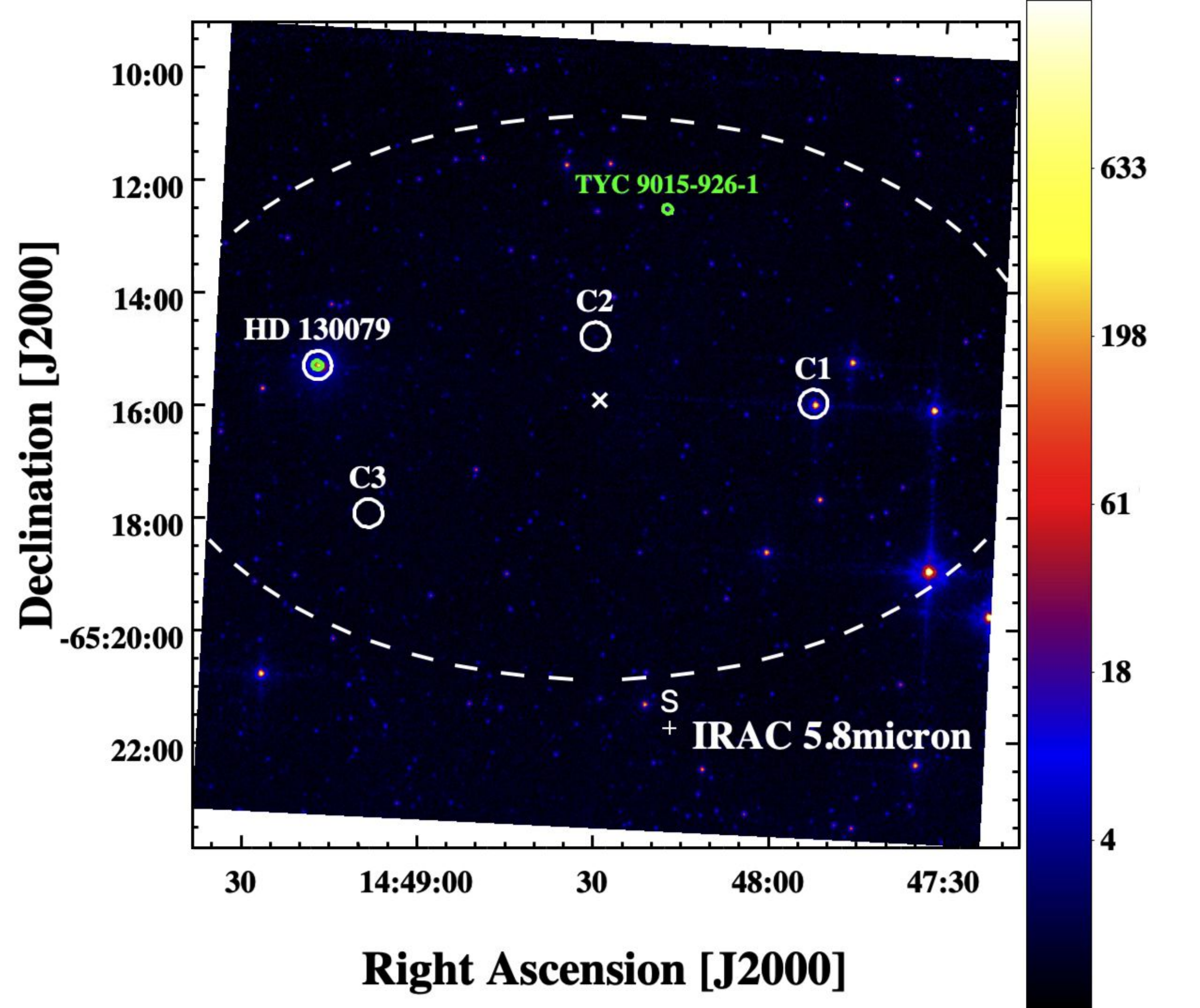}
\includegraphics[width=0.49\textwidth]{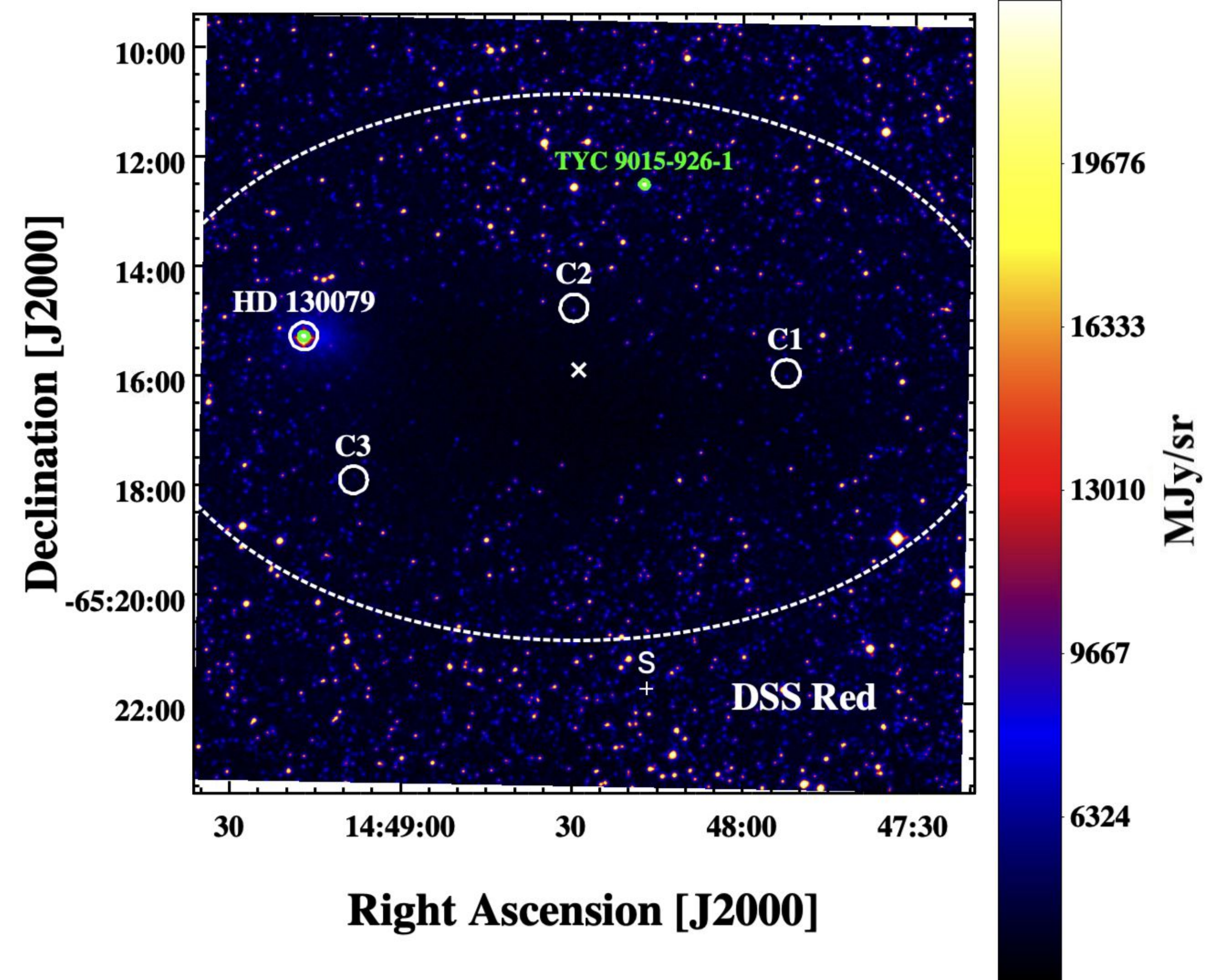}
\includegraphics[width=0.49\textwidth]{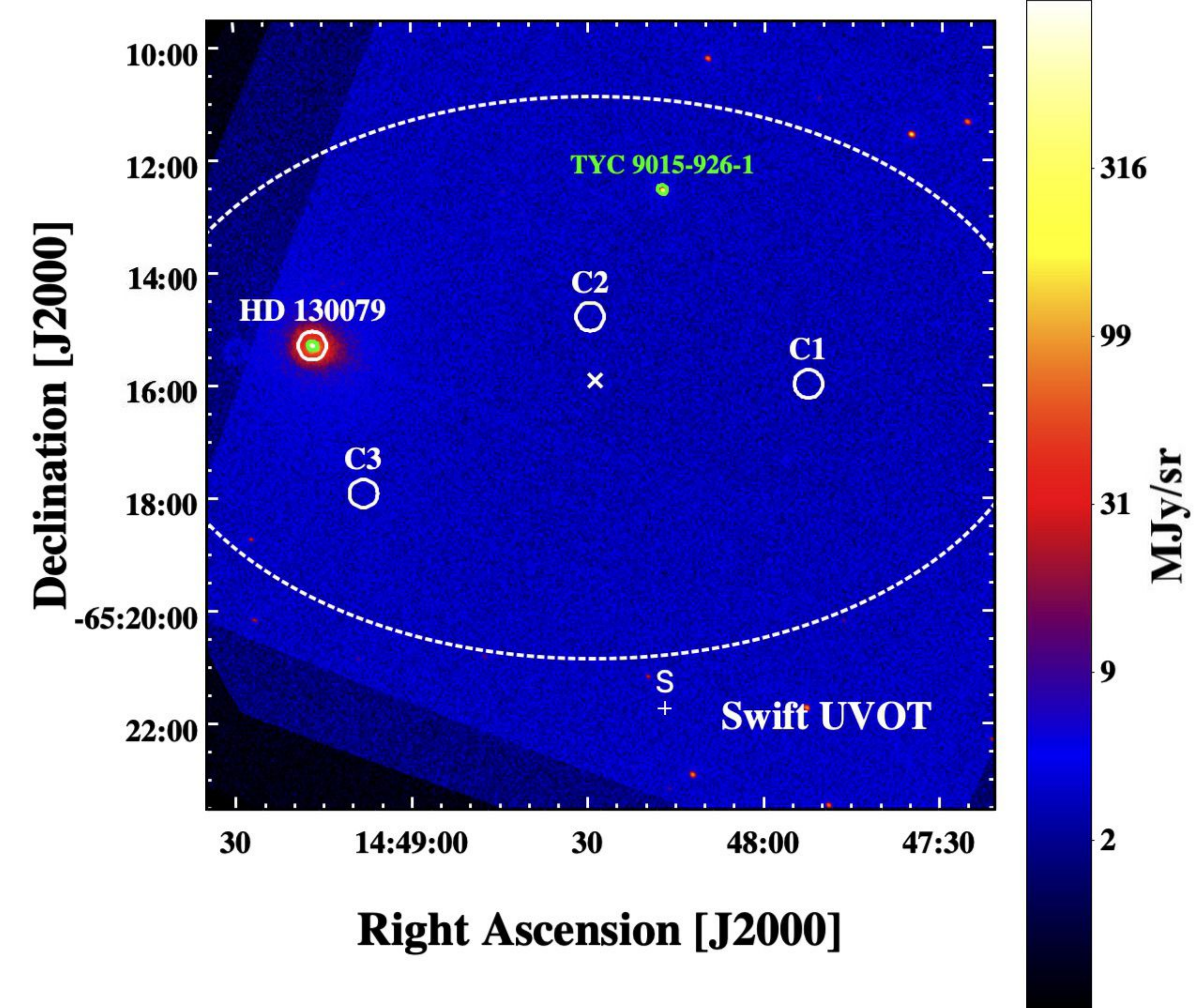}
 \caption{ DC\,314.8--5.1 region as seen at different wavelengths: (top left) Spitzer MIPS 24\,$\mu$m log-scaled intensity mosaic map; (top right) Spitzer IRAC 5.8\,$\mu$m log-scaled intensity mosaic map; (bottom left) DSS red linear scaled image (700\,nm); (bottom right) Swift UVOT M2--2250\AA\, band log-scaled map. In each panel, the white dashed ellipse denotes the globule with the central position marked by a white``x''. The green ellipses mark UVOT detected sources with HD\,130079 marked on the left and TYC 9015-926-1 marked near the northern boundary of the globule. ``C1'' marks the YSO candidate identified by \citet{Whittet07}. ``C2'' and ``C3'' mark the potential YSO candidates identified in this work. The X-ray source detected with Swift-XRT is indicated by ``S'' with a cross.} 
\label{fig:MIPS-UVOT}
\end{figure*}

\section{Multi-wavelength Overview} 
\label{sec:mw_intro}

The DC\,314.8--5.1 dark globule is located approximately 5\,deg below the Galactic plane in the southern constellation Circinus (see Table\,\ref{table:S}). The B9\,V field star HD\,130079, located near the cloud's eastern boundary illuminates a reflection nebula \citep{Whittet07}. The association of HD\,130079 with DC\,314.8--5.1 was established by \citet{vandenBergh75} who identified the presence of the reflection nebula through a survey of southern globules with the Cerro Tololo Observatory. \citet{vandenBergh75} further characterized the host cloud through absorption around the reflection nebulae, determined from the density of field stars method. Later, \citet{Bourke95b} used NH$_3$ observations to determine the physical characteristics (density, temperature, mass) of isolated dark clouds, including DC\,314.8--5.1. 

The parallax value for HD\,130079 in Gaia Early Data Release 3 \citep[EDR3;][]{Gaia21}, is $2.2981 \pm 0.0194$\,mas. \citet{Bailer-Jones21} used the Gaia data and additional analyses to estimate the distance to the star as $431.7^{+3.2}_{-4.3}$\,pc. Using this value as the distance to the cloud, the cloud's $\sim 7^{\prime} \times 5^{\prime}$ radial angular dimensions translate to projected linear sizes of $0.9\,{\rm pc} \times 0.6$\,pc, while the mean atomic hydrogen core number density and the total cloud mass inferred from the extinction characteristics \citep[see][]{Whittet07}, become $\sim 7\times10^3$\,cm$^{-3}$ and $\sim 160 M_{\odot}$, respectively.

\begin{deluxetable*}{crrrrccc}[th!]
\movetableright=0.05cm
\tabletypesize{\scriptsize}
\tablecaption{Source Associations}
\tablewidth{0pt}
\tablehead{
\colhead{Name}& \colhead{RA (J200)} & \colhead{Dec (J200)} &\colhead{Gaia Parallax}& \colhead{Distance} & \colhead{Source Identification} & \colhead{Identification Method} & \colhead{Reference} \\
\colhead{~~}& \colhead{[HH:MM:SS]} & \colhead{[DD:MM:SS]} & \colhead{mas} & \colhead{pc} & \colhead{~~} & \colhead{~~} & \colhead{~~} \\
\colhead{(1)} & \colhead{(2)} & \colhead{(3)} & \colhead{(4)} & \colhead{(5)} &  \colhead{(6)} & \colhead{(7)}   & \colhead{(8)}  
}
\tiny
\startdata
DC\,314.8--5.1 & 14:48:29.00 & --65:15:54.00 & -- & -- & Cloud Center & Visual Extinction Map & \citet{Hartley86} \\
HD\,130079 & 14:49:16.55$^{\dagger}$ & --65:15:18.72$^{\dagger}$ & $2.298\pm 0.019$ & $431.7^{+3.2}_{-4.3}$ & B9V star & Spectral Classification & \citet{Whittet07} \\
S & 14:48:16.72 & --65:21:45.60& -- & -- & Background galaxy & Color-color$^{\ddagger}$ & This Work \\
2RXS & 14:48:33.73 & --65:17:38.90 & -- & -- & Unknown & X-ray & This Work; \citet{Boller16}\\
TYC\,9015-926-1 & 14:48:17.13 & --65:12:33.05 & $2.167\pm 0.015$ & $456.6^{+3.1}_{-3.2}$ & Background star & Photometric Measurement & \citet{Hog00}\\
C1 & 14:47:52.20 & --65:16:01.00 & $0.073\pm 0.096$ & $6670^{+3750}_{-2250}$ & Background star & SED$^{\star}$ & This Work; \citet{Whittet07} \\
C2 & 14:48:29.39 & --65:14:48.52 & $2.262\pm 0.059$ & $436^{+11}_{-11}$ & Class\,III/Field Star & SED$^{\star}$ & This Work \\
C3 & 14:49:07.96 & --65:17:56.41 & $2.441\pm 0.310$ & $442^{+75}_{-53}$ & Field Star & SED$^{\star}$ & This Work 
\vspace{5pt}
\enddata
\tablenotetext{}{{\bf col(1)} --- Object name; 
{\bf col(2)} --- Right Ascension; 
{\bf col(3)} --- Declination;
{\bf col(4)} --- Gaia parallax; 
{\bf col(5)} --- Distance from the Gaia parallax and reported by \citet{Bailer-Jones21}; 
{\bf col(6)} --- Source identification; 
{\bf col(7)} --- Identification method;
{\bf col(8)} --- References for cols(6-7).\\ 
$^{\dagger}$ RA \& Dec for HD\,130079 is based on the Swift-UVOT data reported in this work. 
$^{\star}$ indicates sources selected in this work to have color-color cuts consistent with the Spectral Energy Distribution (SED) of pre-stellar objects; $^{\ddagger}$ source detected in this work by Swift-XRT and identified with the aid of WISE color-color diagrams. }
\label{table:S}
\end{deluxetable*}

\subsection{Planck}
\label{sec:Planck}

The top panel of Figure\,\ref{fig:low-res} presents the Planck map of the DC\,314.8--5.1 region at 353\,GHz; the Planck emission peak is offset by $1^{\prime}.4$ to the east from the nominal center of DC\,314.8--5.1, per Table\,\ref{table:S}.

DC\,314.8--5.1 is listed in the Planck Catalogue of Galactic Cold Clumps \citep[PGCC;][]{Planck16} as PGCC G314.77--5.14. The Planck team created cold residual maps by subtracting a warm component from individual maps (at given frequencies) as described in \citet{Planck11}. As such, cold sources will appear as positive departures, signifying lower temperatures than the surrounding background. The modeling of the cloud on the Planck 857\,GHz cold residual map with an elliptical Gaussian returns FWHMs along the major and minor axes of $8^{\prime}.36\pm 0.34$ and $4^{\prime}.60\pm 0.19$, respectively. Through fitting a modified blackbody, $F_{\nu} \propto B_{\nu}\!(T) \times (\nu/\nu_0)^{\beta}$, to the Planck photometric band fluxes, a dust temperature of $T=14\pm1$\,K and spectral index $\beta = 1.8\pm 0.2$ were derived.

The \citet{Planck16} estimate of the distance to DC\,314.8--5.1 from near-infrared extinction is $400\pm370$\,pc. When combined with the Planck photometry, the resulting mass estimate and mean density for the cloud are $10\pm 14\,M_{\odot}$ and $892\pm544$\,cm$^{-3}$, respectively. For comparison, using the Gaia measured distance, $\sim 432$\,pc, and considering only the error in the Planck flux measurement, we derive a mass of $12.0 \pm 0.8\,M_{\odot}$. Though somewhat unconstrained, these are both below the corresponding estimates by \citet{Hetem88} and \citet{Whittet07}, of 25 and 50\,$M_{\odot}$, respectively.

\subsection{IRAS}
\label{sec:IRAS}

The Point Source Catalog for IRAS has angular resolutions of $0^{\prime}.5$, $0^{\prime}.5$, $1^{\prime}.0$, and $2^{\prime}.0$ for 12, 25, 60 and 100\,$\mu$m, respectively. The catalog has completeness down to levels of 0.4, 0.5, 0.6, and 1.0 Janskys (Jy) at 12, 25, 60, and 100$\mu$\,m, respectively \citep{IRAS}. At the adopted distance of the cloud these correspond to monochromatic luminosities of $2.2\times10^{33}$\,erg\,s$^{-1}$ ($\sim0.58 L_{\odot}$) at 12\,$\mu$\,m, $1.3\times10^{33}$\,erg\,s$^{-1}$ ($\sim0.34 L_{\odot}$) at 25\,$\mu$\,m, and $6.7\times10^{32}$\,erg\,s$^{-1}$ ($\sim0.18 L_{\odot}$) at 60 and 100\,$\mu$\,m \citep{IRAS}.

Three point sources were identified in the IRAS Point Source Catalog coincident within 10$^\prime$ radius around the center of DC\,314.8--5.1, namely IRAS\,14451--6502, IRAS\,14437-6503, and IRAS\,14433-6506; and no sources were identified in the Faint Source Catalog \citep{IRAS}. Source IRAS\,14451--6502 is associated with HD\,130079, with high-quality detections in the first three bands and moderate quality detection in the 100\,$\mu$m band, see Table\,\ref{table:S}. IRAS\,14437-6503 was identified in \citet{Bourke95a}, along with HD\,130079, to be associated with DC\,314.8--5.1. However, IRAS\,14437-6503 corresponds to the Gaia DR3 source 5849039334515066624 with a Gaia measured parallax of $0.073\pm0.0967$ corresponding to a \citet{Bailer-Jones21} measured distance in the range $4.4-10.4$\,kpc and as such is not physically related to the cloud. IRAS\,14433-6506 is located at the outskirts of the cloud and associated with the Gaia DR3 source 5849037788326819072 with a Gaia measured parallax of $0.866\pm0.027$ corresponding to a \citet{Bailer-Jones21} measured distance range of $\simeq 1.1$\,kpc and as such is determined to not be associated with DC\,314.8--5.1. 

The bottom panel of Figure\,\ref{fig:low-res} presents the  60\,$\mu$m IRAS map of DC\,314.8--5.1, with an angular resolution of $\sim 1^\prime$ \citep{Wheelock94}. The far-infrared intensity is shifted to the east of center by $\sim 3^{\prime}.5$ likely due to heating by HD\,130079.

\subsection{ROSAT}
\label{sec:ROSAT}

A weak X-ray point source is present in the Second ROSAT all-sky survey (2RXS), with the survey having an effective angular resolution of $1^{\prime}.8$ \citep{Boller16}, suggesting that DC\,314.8--5.1 could be an X-ray emitter. The source position for ROSAT J144833.7--651738 is $1^{\prime}.8$ to the south of the cloud center (see Figure\,\ref{fig:ROSAT-XRT} in Appendix\,\ref{A:ROSAT}), but still contained within the cloud's boundary, per Table\,\ref{table:S} denoted as ``2RXS''. The low photon count of $8\pm4$\,cts in the  ROSAT Position Sensitive Proportional Counters (PSPC) $0.1-2.4$\,keV band is insufficient for any meaningful spectral modeling.

\subsection{Spitzer}
\label{sec:Spitzer}

The Spitzer Space Telescope observational data for this work, obtained from the NASA/IPAC Infrared Science (IRSA) archive, were originally acquired with the the Infrared Array Camera \citep[IRAC;][]{Fazio04} and the Multiband Imaging Photometer \citep[MIPS;][Proposal ID\,50039; P.I.: D.~Whittet]{Rieke04}. DC\,314.8--5.1 was observed for a total of 6 hours in 2008 October in five infrared bands: 3.6, 4.5, 5.8, and 8.0\,$\mu$m with IRAC, and 24\,$\mu$m with MIPS, with angular resolutions of $\sim 2^{\prime\prime}$ and $\sim 6^{\prime\prime}$, respectively.

Due to the presence of many bright sources within the field, we performed artifact correction utilizing the IRAC artifact mitigation tool, by following a procedure similar to that in the Spitzer Data Cookbook\footnote{\label{note1}\url{https://irsa.ipac.caltech.edu/data/SPITZER/docs/}}, and additional tools listed within, to produce mosaic maps in each band. The resulting 5.8\,$\mu$m IRAC image is shown in the upper right panel of Figure\,\ref{fig:MIPS-UVOT}, while all four IRAC images are shown in Appendix\,\ref{A:MIR-images} as Figure\,\ref{fig:IRAC}. Reduction of the MIPS data similarly followed recipe 22 in the Spitzer Data Cookbook using MOPEX \citep{Mopex}. The resulting MIPS 24\,$\mu$m map of the region, is shown in the top left panel of Figure\,\ref{fig:MIPS-UVOT}. 

\begin{figure*}[th!]
\centering
\includegraphics[width=0.75\textwidth]{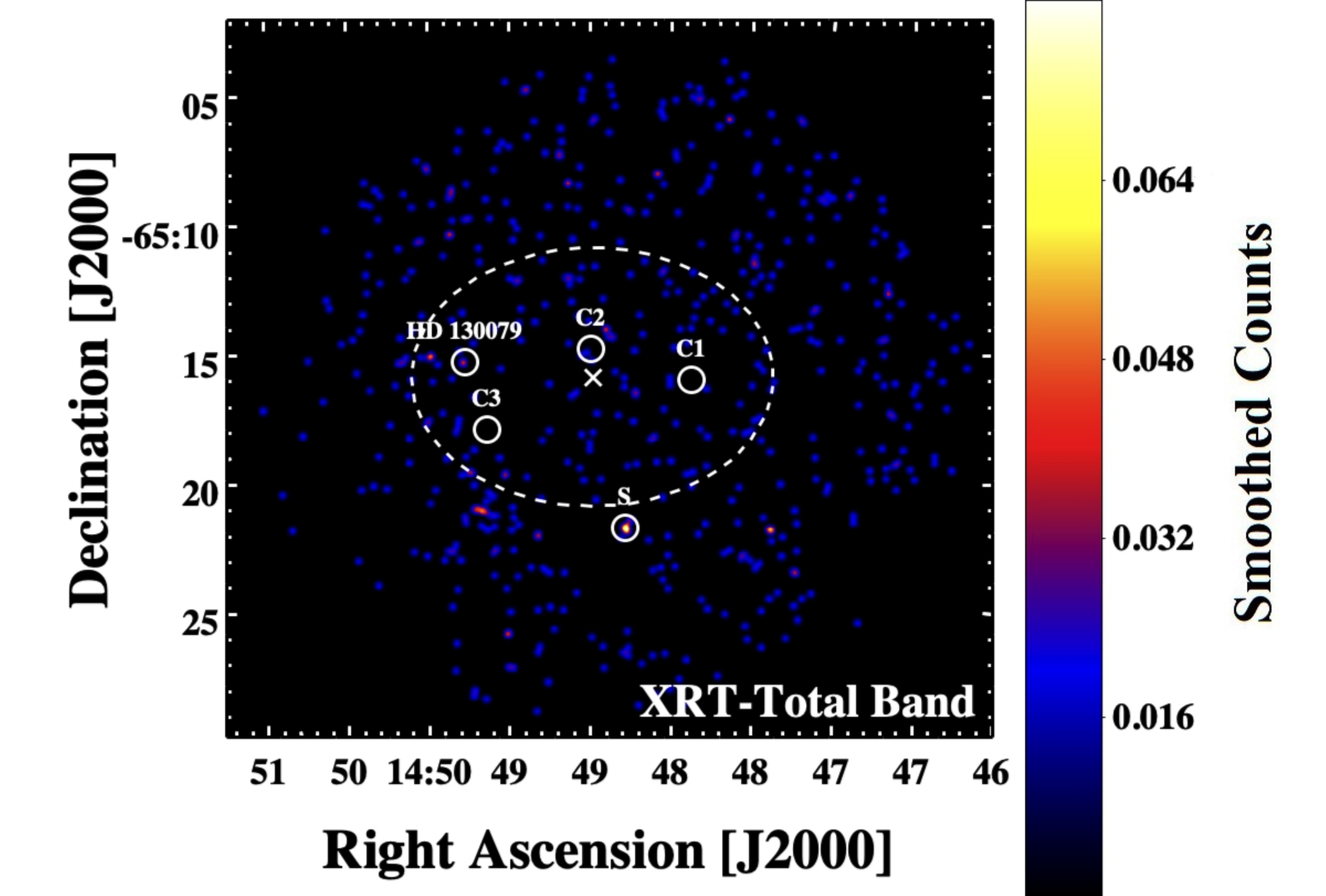}\\
\includegraphics[width=0.49\textwidth]{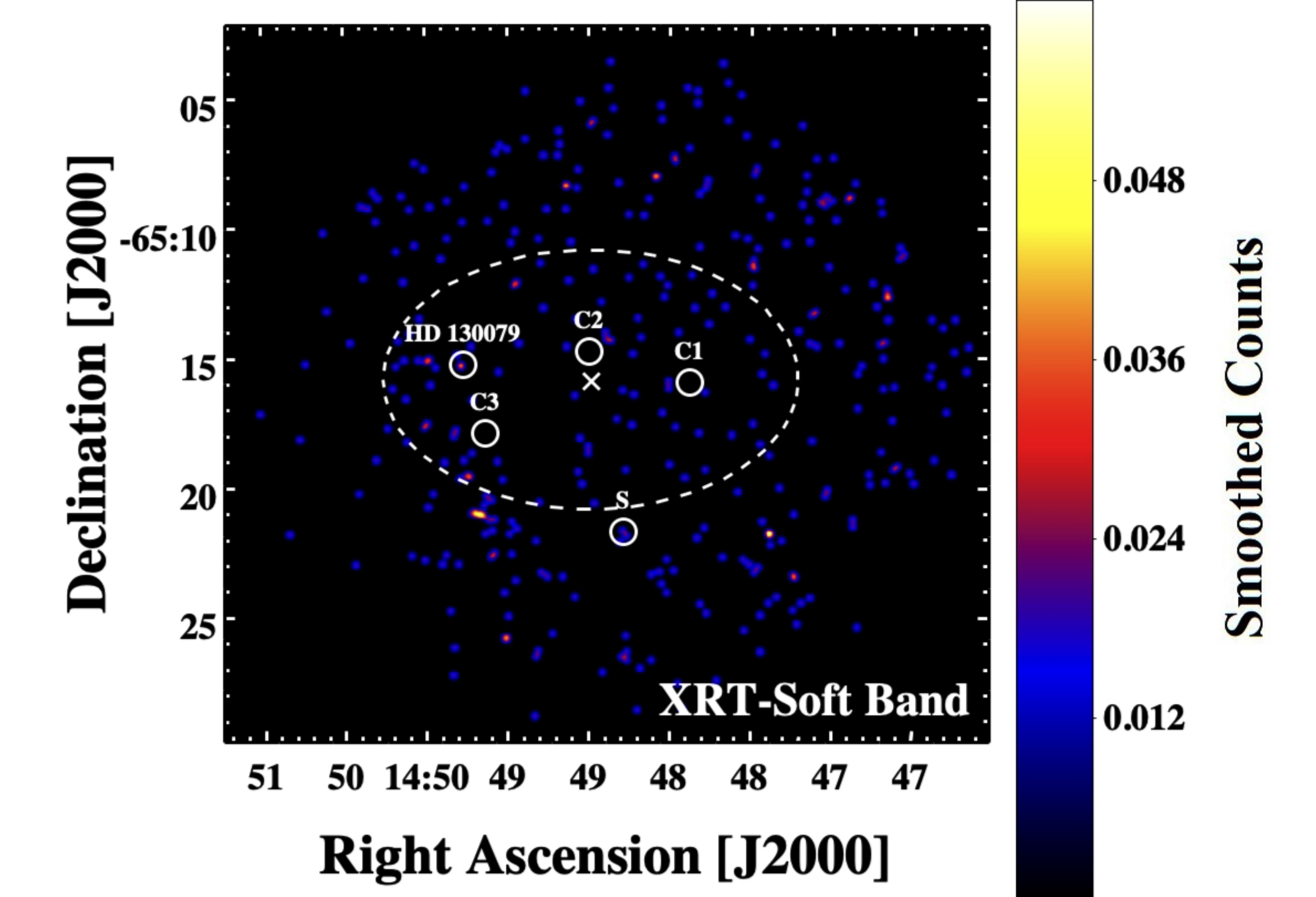}\hspace{0.1cm}
\includegraphics[width=0.49\textwidth]{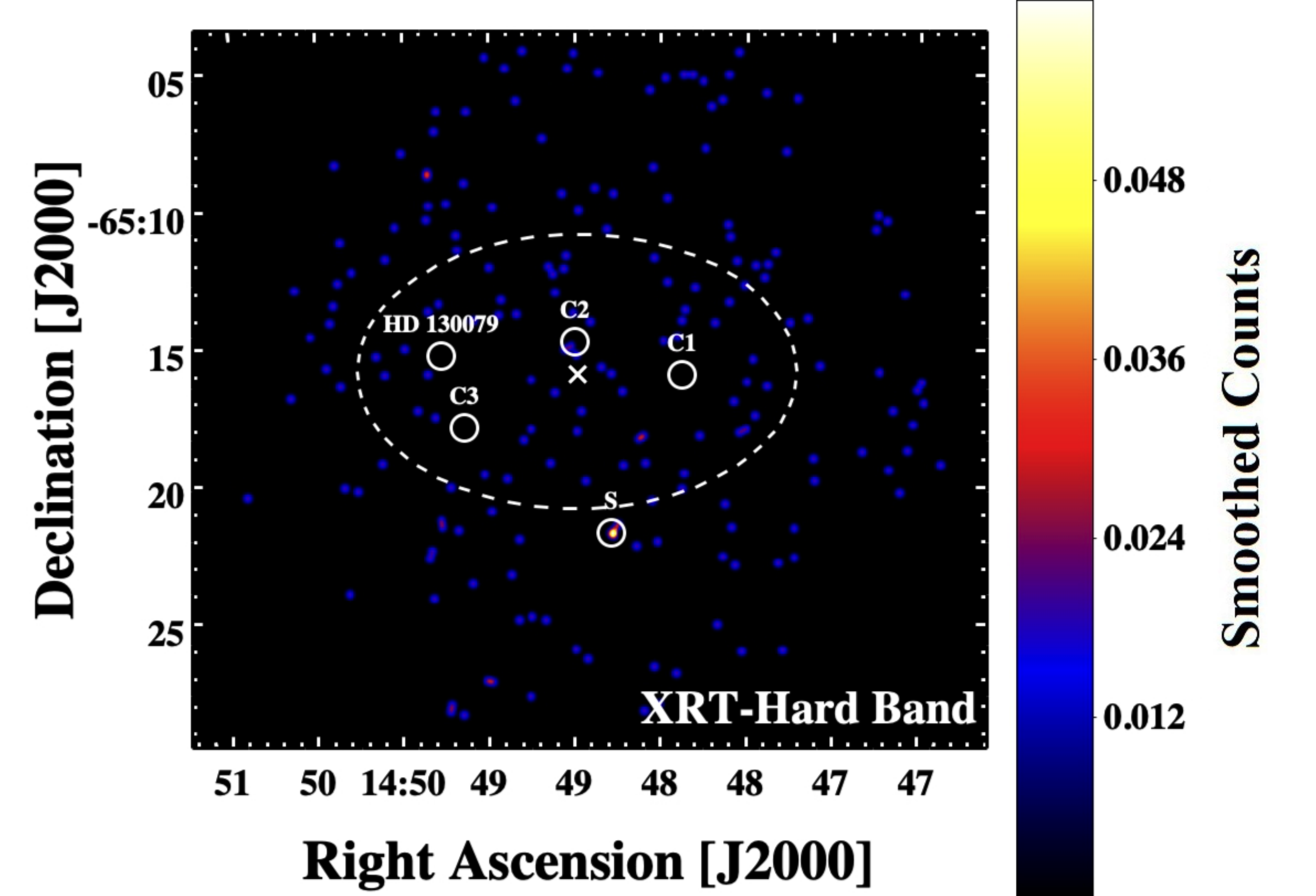}
\caption{
(top) Full-band 0.3--10\,keV Swift XRT image of the  DC\,314.8--5.1 region, smoothed with a Gaussian of radius 6\,pixels. HD\,130079 field star is marked in the left of each image. ``C1'' marks the YSO candidate identified by \citet{Whittet07}. ``C2'' and ``C3'' mark the potential YSO candidates identified in this work. Colorbar indicates the linear intensity scale for the smoothed (averaged) counts. (bottom left) Soft-band 0.3--2.0\,keV and (bottom right) hard-band 2--10\,keV. }
\label{fig:XRT}
\end{figure*}

\subsection{WISE}
\label{sec:WISE}

WISE was an all-sky survey mission covering the entirety of the northern and southern hemispheres. The telescope obtained measurements in four bands (W1--W4) centered at 3.4, 4.6, 12, and 22\,$\mu$m with resolutions of $6^{\prime\prime}.1$, $6^{\prime\prime}.4$, $6^{\prime\prime}.5$, and $12^{\prime\prime}$, respectively \citep{Wright10}. 

\subsection{DSS}
\label{sec:DSS}

The DC\,314.8--5.1 spatial extent was delineated by \citet{Whittet07} based on the opacity spatial distribution seen in the Digitized Sky Survey (DSS) image,  shown here as Figure\,\ref{fig:MIPS-UVOT}, bottom left panel. We note that the maximum cloud core visual extinction, $A_v$, according to \citeauthor{Whittet07} is $\gtrsim 8.5$\,mag through the center of the core and decreasing towards the outer regions.

\subsection{Swift XRT \& UVOT}
\label{sec:Swift}

Swift Target of Opportunity (ToO) observations of DC\,314.8--5.1 were obtained with the Swift X-Ray Telescope (XRT) instrument \citep{Burrows00}, as well as the Ultra-violet Optical Telescope \citep[UVOT;][]{Roming05} filter of the day, in this instance the UVM2--2250\AA\, band (Proposal ID: 16282, Requester: E.~Kosmaczewski). The target was observed for 3\,ks on 2021 September 26. Swift XRT observations were taken in photon counting mode. Three Swift XRT images were produced utilizing the Swift XRT data products generator: the entire spectral range from 0.3--10\,keV; the soft band from 0.3--2.0\,keV; and, the hard band from 2--10\,keV. The procedure for image creation followed \citet{Evans20}.

The resulting UVOT M2--2250\AA\, band image of the  DC\,314.8--5.1 region is shown in the bottom right panel of Figure\,\ref{fig:MIPS-UVOT}. The XRT maps of the cloud are shown as Figure\,\ref{fig:XRT}, including the full-band image from 0.3--10\,kev (top panel), the soft-band image from 0.3--2.0\,keV (bottom-left), and the hard-band image from 2--10\,keV (bottom-right). Images have been smoothed to aid in visualization \citep{Joye03}, utilizing a Gaussian profile with a radius of 6\,px, and $\sigma =3 $\,px.

Source detection was performed for each of the three Swift XRT Point Source Catalogue (2SXPS) energy bands (i.e., 0.3--1.0\,keV, 1--2\,keV, 2--10\,keV), as well as for the total energy range (0.3--10\,keV), following \citet{Evans20}. No sources were found within the individual narrow bands, and only one low-significance source, denoted hereafter as ``S'', was detected within the total energy range. The location of the source S, see Table\,\ref{table:S}, places it at, or just beyond, the periphery of the cloud. The error in the position measurement is $6^{\prime\prime}.9$, and the XRT source off-axis angle is $5^{\prime}.8$. The source was detected with $C=8$\,cts including background counts, with average background $B \simeq 0.5$\,cts. The corresponding errors, calculated according to \citet{Gehrels86}, as appropriate in the regime of a very low photon statistics \citep[see][]{Evans14}, are
\begin{equation}
    \sigma_C = 1.0 + \sqrt{C+0.75} \simeq 4.0
\end{equation}
and analogously $\sigma_B \simeq 2.1$, leading to a signal-to-noise (SNR) for the detection of
\begin{equation}
    {\rm SNR} = \frac{C-B}{\sqrt{\sigma_C^2 +\sigma_B^2}} \simeq 1.6 \, .
\end{equation}

The source ``S'' can be seen at a low level in the hard-band image but it is not distinguishable in the soft-band image (see bottom panels in Figure\,\ref{fig:XRT}), which may indicate a hard X-ray spectrum for the source. However, the low photon numbers prevent further characterization of its spectrum. An additional peak can be seen in the lower left just outside the cloud core in the total (top panel) and soft band (bottom left panel) in Figure\,\ref{fig:XRT}. However, this peak fails to meet a $\sigma\ge1$ and as such is not discussed further here.

The Swift UVOT source extraction was performed utilizing the standard ``uvotdetect'' routine \citep{Roming05}. We detected a total of 38 sources to a detection threshold of $5 \sigma$. Only two sources (see the bottom left panel of Figure\,\ref{fig:MIPS-UVOT}) are detected within the extent of the cloud, with the brightest being HD\,130079. The second source is a star, TYC\,9015-926-1, in the northern region of the cloud, see Table\,\ref{table:S}. It has a Gaia measured parallax of $2.167\pm 0.015$\,mas, corresponding to a \citet{Bailer-Jones21} distance range of $453.4-459.7$\,pc. Therefore, this object is a (somewhat) nearby star located behind the cloud.

\section{Identification of YSOs} 
\label{sec:YSO}

The evolutionary progression of the infrared emission of YSOs is such that the wavelength of the blackbody peak emission migrates towards the near-infrared as the YSO ages, while the far-infrared excess (due to the surrounding dusty disks and/or envelopes) decreases \citep[see, e.g.,][]{Andre00, Greene94}. This results in the population of the youngest YSOs, i.e. Class 0 sources, emitting almost exclusively in the sub-mm/far-infrared range. On the other hand, Class I--III sources, which emit efficiently at shorter wavelengths, if present, should manifest in the analyzed mid-infrared surveys \citep[see in this context][]{Gutermuth09, Evans09}. Class I sources are deeply-embedded protostars with infalling, dense envelopes, characterized by a rising or flat mid-infrared spectrum. Class\,II sources denote YSOs that are pre-Main Sequence Stars with gas-rich optically-thick disks and on-going accretion onto the central star, and a decreasing MIR spectrum. Finally, Class III YSOs have gas-poor disks and very little infrared excess due to dust, and are notoriously difficult to separate from young Main-Sequence stars. We also discuss here the so-called ``transition disk'' objects which are YSOs without an inner disk but containing an optically thick outer disk \citep{Andre00}. 

\subsection{Class 0 Source Limits}

The Point Source Catalog for IRAS identified no candidate sources within our system down to luminosity levels $6.7\times10^{32}$\,erg\,s$^{-1}$ ($\sim 0.18 L_\odot$) for 60 and 100\,$\mu$\,m, see Section\,\ref{sec:IRAS}. This luminosity level indicated that our study is sensitive to YSO Class\,0 sources down to core masses $\sim 0.1M_\odot$ \citep{Dunham12}. The lack of any source detections associated with the cloud, with the exception of HD\,130079, strongly suggests the absence of YSO Class\,0 sources within the system.

\subsection{Spitzer IRAC \& 2MASS Source Examinations}
\label{sec:IRAC-YSO}

The Spitzer IRAC mapping data, for the observed frame time of 12 seconds, effectively probes down to flux levels of 52\,$\mu$Jy at 8\,$\mu$m, and 6.1\,$\mu$Jy at 3.6\,$\mu$m with a spatial resolution of $\sim 2\arcsec$ \citep{Fazio04}. At the distance of the cloud (432\,pc), these limits correspond to monochromatic luminosities of $\simeq 4.4 \times 10^{29}$\,erg\,s$^{-1}$ and $\simeq 1.2 \times 10^{29}$\,erg\,s$^{-1}$, respectively. The observed 3.6\,$\mu$m range (3.1--3.9\,$\mu$m), in particular, is rather close to the peak of the blackbody emission component in Class\,I--III sources, and as such the latter value should serve as a good proxy for the limiting luminosity of YSOs candidates, with the bolometric correction of the order of a few at most \citep[see, e.g.,][]{Lada87}. In other words, in the Spitzer IRAC mapping data, we are sensitive to YSO Class\,I--III luminosities as low as $\sim 10^{-4} L_{\odot}$, so that any young star with a core mass down to $0.01 M_{\odot}$ \citep[see][]{Dunham12}, should easily be detected.

We performed a search with a radius of $5^{\prime}$ around the central position of the cloud, see Table\,\ref{table:S}, with the Spitzer Enhanced Imaging Products (SEIP) source list in order to identify potential YSOs. We restricted our sample by a signal-to-noise SNR\,$>5$ in all four IRAC bands, excluding unresolved extended sources and excluding sources with only upper limits in any band (sources detected in only some bands are considered in the follow-up selection). This returned a total of 1,319 sources within the sampled region.

First, we applied the color criteria from \citet{Gutermuth09}, Appendix A.1 therein, to the sample of 1,319 sources. This removed 132 star-forming galaxies (SFGs) and 256 active galactic nuclei (AGN) resulting in a sample of 924 potential YSOs. \emph{None} of these sources met the criteria to be identified as a Class\,I or Class\,II YSO as defined in \citet{Gutermuth09}.

Second, we investigated sources with lower significant detections, following the cuts presented in \citet{Winston19}, Appendix A.2 (Equations 17--20). Specifically, those sources that are lacking robust (SNR\,$<5$) detections in IRAC 5.8\,$\mu$m or IRAC 8.0\,$\mu$m, but still show SNR\,$>5$ detections in IRAC 3.6\,$\mu$m and IRAC 4.5\,$\mu$m, with the requirement that they also have significant ($\sigma<0.1$\,mag) detections in 2MASS bands H and K$_s$. However, we find no sources within this sample that meet the color criteria needed for a YSO detection as defined in \citet{Winston19}.

Further, we searched for deeply-embedded protostars, in the so-called ``Phase 3'' cuts adopted by \citet{Gutermuth09}, Appendix A.3 therein. We included sources from SEIP that lack detections in IRAC 5.8\,$\mu$m or IRAC 8.0\,$\mu$m bands, are bright in the MIPS 24\,$\mu$m band, and have strong (SNR$>5$) detections in IRAC 3.6\,$\mu$m and IRAC 4.5\,$\mu$m. Our selection returned 164 sources, including some previously flagged as AGN based on the IRAC color cuts. However, only three sources met the MIPS 24\,$\mu$m band brightness criteria of $[24]<7$\,mag. None of these three sources satisfied the remaining criteria to be identified as a YSO, and as such this selection returned no candidate sources.

Finally, we comment here on the remaining sources not classified in the first \citet{Gutermuth09} cuts adopted here, see the right panel in the Appendix\,\ref{A:IRAC-cuts} Figure\,\ref{fig:IRAC-cuts}. Sources that fall in this range are often consistent with Class\,III sources (see \citet{Dunham15, Anderson22} for further discussion). However, these regions of infrared colors are heavily contaminated by AGB type background stars. \citet{Dunham15} estimated contamination in the Class\,III type sources by background stars ranges from $25-90\%$ in their sample. In order to disentangle background stars from true Class\,III sources, we further inspected these remaining 923 sources with Gaia, below.

 \subsection{Gaia Parallax Measurements}
\label{sec:Gaia}

We inspected the remaining 923 IRAC sources unidentified by the \citet{Gutermuth09} cuts, which are likely background stars or Class\,III candidates (see Section\,\ref{sec:IRAC-YSO}), with the Gaia source catalog \citep{Gaia21}. Gaia parallaxes provide precise measurements with a spatial resolution of $0^{\prime\prime}.4$ and so are capable of separating individual objects even when clustered on the sky.

The source ``C1'', not identified above as a bona fide YSO, but previously identified as a potential candidate by \citet{Whittet07}, has a Gaia measured parallax of $0.073\pm 0.096$\,mas. The \citet{Bailer-Jones21} catalog marks the distance to this star as $6.67^{+3.75}_{-2.25}$\,kpc, which is far beyond DC\,314.8--5.1.

We looked at a sample of Gaia sources within the same region inspected by IRAC, out to a radius of $5^{\prime}$ around the central position of the cloud, see Table\,\ref{table:S}. We additionally constrained the list of Gaia sources to those having a parallax measurement (within the error bounds) coinciding with the parallax for HD\,130079, i.e. $2.2981\pm0.0194$. To cross-check with the 923 IRAC sources, we investigated each IRAC source for any ``good'' Gaia source within the IRAC spatial resolution of $\sim2^{\prime\prime}$ \citep{Fazio04}. This resulted in a sample of 27 potential Class\,III/Field Star sources. We further cross checked this sample with the the \citet{Bailer-Jones21} catalog for measured distances consistent with HD\,130079 ($\sim 427-435$\,pc). We identified 2 sources, SSTSL2 J144829.39-651448.5 and SSTSL2 J144907.95-651756.4, corresponding to the Gaia DR3 sources, id:\,35849036757534689536,  and 5849041288680373504, with appropriate distance measurements, denoted hereafter as ``C2'' and ``C3'', per Table\,\ref{table:S}.

\subsection{Pre-Main Sequence Stars in Swift XRT Data}
\label{sec:PMS}

PMSs are established X-ray emitters, with corresponding X-ray luminosities 10--10,000 times above the levels characterizing the old Galactic disk population \citep[e.g.,][]{Preibisch05,Tsuboi14}. They are routinely detected with the Chandra X-ray Observatory in molecular clouds because their keV photons penetrate heavy extinction \citep[e.g.,][]{Wang09,Kuhn10}. The bright members of PMS populations revealed by such studies are typically well modeled assuming a plasma in collisional ionization equilibrium, using the Astrophysical Plasma Emission Code \citep[APEC;][]{Smith01}, with temperatures of the order of a few-to-several keV, low metal abundances, and 0.5-10\,keV luminosities of the order of $10^{30}$\,erg\,s$^{-1}$. 

Here, we compare the expected X-ray levels of PMSs in DC\,314.8--5.1 with X-ray luminosity of the source ``S'' detected in the Swift XRT pointing. We calculate the Galactic hydrogen column density in the direction of DC\,314.8--5.1, utilizing the NHtot tool provided through HESEARC, \citep{Bekhti16}. Given that the source is only at a distance of 432\,pc, the resulting values of $N_{\rm H,\,Gal} \simeq 3.2 \times 10^{21}$\,cm$^{-2}$ is likely an overestimation for the real column density along the line of sight. Nonetheless, in all the flux estimates below we conservatively adopt the value $N_{\rm H,\,Gal} \simeq 3 \times 10^{21}$\,cm$^{-2}$. 

In addition to the Galactic diffuse ISM fraction, we take into account the intrinsic absorption within the cloud. For this, assuming the cloud's mean gas density of $10^4$\,cm$^{-3}$ and a spatial scale of 0.3\,pc, the corresponding column density is estimated at the level of $N_{\rm H,\,int} \simeq 10^{22}$\,cm$^{-2}$. Again, this should be considered as an upper limit for the intrinsic absorption value.

For the APEC spectral model with temperature $kT = 2$\,keV, metallicity $Z = 0.2\,Z_{\odot}$, and the absorbing column densities as estimated above, the $0.3-10$\,keV XRT observed count rate integrated over all solid angle, $4.1^{+1.8}_{-1.4} \times 10^{-3}$\,cts\,s$^{-1}$, corresponds to an unabsorbed flux of $F_{\rm 0.3-10\,keV} \simeq 4^{+1.7}_{-1.4} \times 10^{-13}$\,erg\,cm$^{-2}$\,s$^{-1}$. The isotropic intrinsic luminosity of $L_{\rm 0.3-10\,keV} \simeq 0.9^{+0.4}_{-0.3} \times 10^{31}$\,erg\,s$^{-1}$ would be $2\times 10^{-3} L_{\odot}$ if at the distance of DC\,314.8--5.1.  However, by decreasing the intrinsic column density down to $N_{\rm H,\,int} \simeq 10^{21}$\,cm$^{-2}$, as would be a more appropriate estimate at the far outskirts of DC\,314.8--5.1, the implied intrinsic luminosity decreases to $L_{\rm 0.3-10\,keV} \simeq 4.9^{+2.1}_{-1.7} \times 10^{30}$\,erg\,s$^{-1}$\,$\sim 10^{-3} L_{\odot}$. 

For comparison, main-sequence high-mass stars are established sources of soft X-ray emission, characterized by an approximately linear luminosity scaling with bolometric photospheric emission, $\log (L_X/L_{\rm bol}) \simeq -7$ \citep{Gudel09}. Yet PMSs are relatively brighter in X-rays, especially in the low-mass range. Indeed, the recent analysis by \citet{Getman22} showed that, for PMS systems with masses $0.1-2 \, M_{\odot}$, the ratio  $\log (L_X/L_{\rm bol})$ can be found in a broad range from $\lesssim -4$ up to $\gtrsim -2$, with the median just below the saturation level\footnote{In cool young stars characterized by rapid rotation, the coronal X-ray luminosities powered by magnetic activity, is observed to saturate at the  $\log (L_X/L_{\rm bol}) = -3$ level \citep{Vilhu87,Stauffer94}.} of --3. This ratio, however, falls rapidly above $2 \, M_{\odot}$. 

During the first few Myr, the X-ray luminosity of PMSs appear approximately constant, declining with time at later evolutionary stages, and again more rapidly as stellar mass increases \citep{Getman22}. Only a small fraction of the $<2 M_{\odot}$ systems appears brighter than $3 \times 10^{30}$\,erg\,s$^{-1}$ in X-rays, and those cases are believed to represent super- and mega-flaring states \citep[e.g.,][]{Getman21}. More than $75\%$ of the more massive ($2-100\, M_{\odot}$) systems, on the other hand, exceed $3 \times 10^{30}$\,erg\,s$^{-1}$. As such, at the Swift-XRT luminosity level of $L_{\rm 0.3-10\,keV} \simeq 4.9^{+2.1}_{-1.7} \times 10^{30}$\,erg\,s$^{-1}$, we are unlikely to detect single PMSs, other than the brightest super/mega flaring sources.

The search for possible infrared counterparts for the Swift XRT source ``S'', returns two WISE/2MASS sources, with $\sim 8^{\prime\prime}-9^{\prime\prime}$ separations from the source ``S'' position. The WISE colors for J144815.66-652141.9 (W1--W2\,=\,--0.62 and W2--W3\,$\geq$\,4.18) indicate that the object could be a background luminous infrared galaxy (LIRG), or a starburst galaxy. The WISE colors of J144818.35-652144.7 (W1--W2\,=\,--0.31 and W2--W3\,$\geq$\,2.37), on the other hand, are consistent with a regular star-forming galaxy \citep[see][]{Wright10}. One of these two sources is a likely counterpart of the Swift XRT source ``S'', neither of which represent a viable PMS candidate.

Furthermore, ROSAT J144833.7–651738 detected with a 448.83\,s exposure by ROSAT, discussed above in Section\,\ref{sec:ROSAT}, is not observed with the 3\,ks observation by Swift-XRT, see Figure\,\ref{fig:XRT}. A comparison of the ROSAT and Swift-XRT maps is shown in Figure\,\ref{fig:ROSAT-XRT} of Appendix\,\ref{A:ROSAT}. This lack of a detection may indicate that ROSAT J144833.7–651738 is an artifact of the 2RXS analysis, or potentially a transient/variable source.

\section{Discussion} 
\label{sec:discussion}

YSOs can be separated from background stars due to the presence of infrared excesses, primarily in the 1--30\,$\mu$m range \citep{Evans09}. As such, if present and related to the cloud, YSOs may appear as optical/infrared point sources for which parallax distances should be similar to the distance of DC\,314.8--5.1. In this context, we investigated the point sources located within the spatial extent of DC\,314.8--5.1 which survived the selection cuts applied following \citet{Winston19} and \citet{Gutermuth09} and with appropriate Gaia parallax and \citet{Bailer-Jones21} distances. Two sources were identified in this way as potential Class\,III candidates: ``C2'' and ``C3.''

\citet{Dunham15} proposed the $[3.6]-[24]\le 1.5$ color values for separating likely Class\,III sources from AGB field stars. Utilizing this cut, and the $3\sigma$ upper limit for the 24\,$\mu$m fluxes, we found color values of $-0.66$ and $2.35$ for C2 and C3, respectively. We can, however, rule out the likelihood of ``C3'' being a Class\,III YSO based on its location near the outer edge of the core region. This is because, on the outskirts of the cloud, we expect a lower level of extinction $\sim 2-3$ (see \citet{Kosmaczewski22, Whittet07}), and so for a Class\,III evolutionary stage source, we would expect to see some evidence of an optical/near-ir reflection nebula \citep{Connelley07, vandenBergh75}. The lack of a detectable reflection nebula for ``C2,'' on the other hand, is unsurprising as ``C2'' is located near the central region of the core with extinction levels $>8.5$\,mag \citep{Whittet07}. Yet this region is also the coldest region (see top panel Figure\,\ref{fig:low-res}) with a Planck measured temperature of $\sim 15$\,K. The presence of a Class\,III source would be expected to produce significant heating of the dust surrounding it, and that is not seen in DC\,314.8--5.1 using available observations \citep{Strom75}. For these reasons, we consider the identification of ``C2'' as a Class\,III YSO to be unlikely. However, detailed spectral modeling combined with deeper X-ray measurements would be necessary to substantiate this claim \citep[see][]{Dunham15}.

The lack of any robust YSO detections further supports the pre-stellar state of DC\,314.8--5.1, as discussed in \citet{Whittet07} and \citet{Kosmaczewski22}. However, younger YSOs (Class\,0) and sources that are still heavily embedded within their cores may not be detectable by mid-infrared excesses \citep{Evans09,karska18}. In order to exclude the presence of such objects deep far-infrared, CO, and/or X-ray observations are needed \citep{Grosso00}.

The short Swift-XRT exposure we have obtained is sensitive to sources within the cloud down to an unabsorbed 0.5--10\,keV luminosity level of $\lesssim 10^{31}$\,erg\,s$^{-1}$. Given this level, only the brightest PMSs could be detected and, among the low-mass ($<2 M_{\odot}$) systems, only young flaring objects would be seen. A much deeper X-ray imaging observation would be needed to constrain the potential PMS population in DC\,314.8--5.1 \citep{Kuhn10,Getman22}.

\begin{figure}[th!]
    \centering
\includegraphics[width=0.47\textwidth]{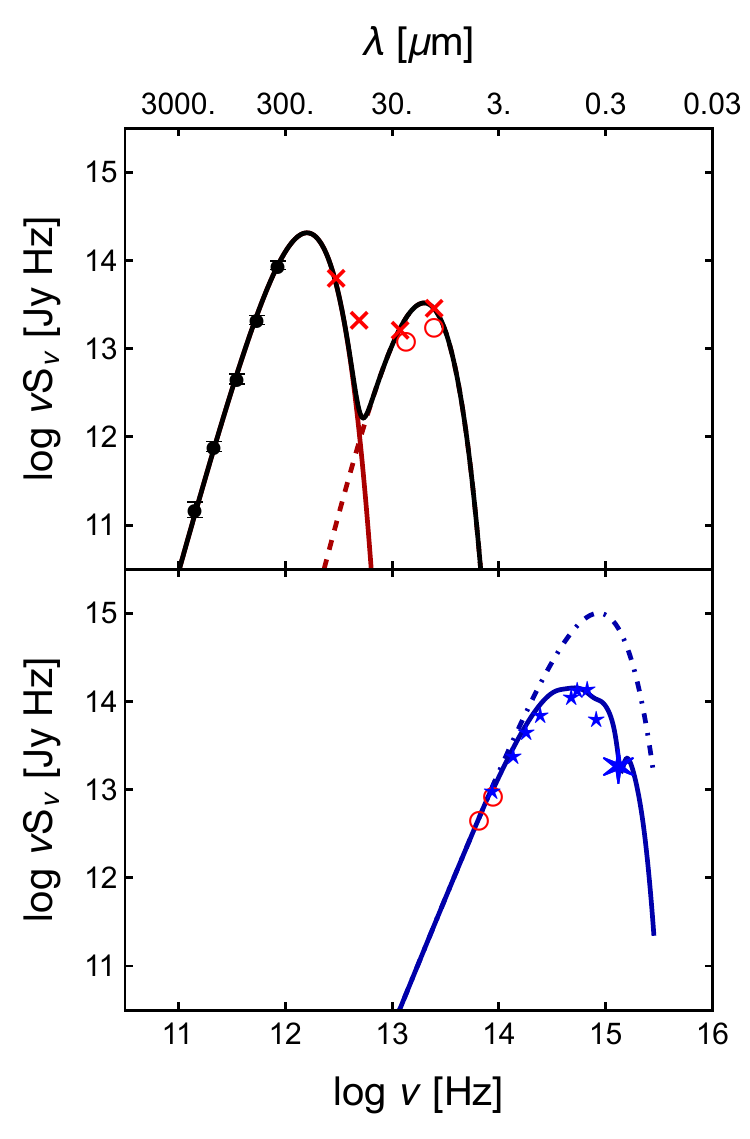}
    \caption{(top) SED of the DC\,314.8--5.1 system, based on observations with Planck (filled black circles), IRAS (red crosses), and WISE (open red circles). Dark red solid and dashed curves represent modified blackbody models for the emission of cold (14\,K) and warm (160\,K) gas within or on the surface of the cloud, respectively; black solid curve denotes the superposition of the two. (bottom) SED of HD\,130079 from ground-based telescopes and Gaia survey (small blue stars), WISE (open red circles) and finally with the Swift UVOT (big blue star). Dark blue dot-dashed curve corresponds to the intrinsic emission of the field star HD\,130079, modeled as a blackbody with the temperature 10,500\,K and the total luminosity of $3 \times 10^{35}$\,erg\,s$^{-1}$; dark blue solid curve illustrates this intrinsic emission subjected to the interstellar reddening. See \S\,\ref{sec:discussion} for description.} 
    \label{fig:SED}
 \end{figure}

The spectral energy distribution (SED) of DC\,314.8--5.1, within the spectral range from microwaves up to UV, is composed of three main components: the thermal emission of the dominant cold dust, the emission of a warm dust photo-ionized and heated by the field star, HD\,130079, and the HD\,130079 photospheric emission itself. These are presented in Figure\,\ref{fig:SED}. 

We consider the cloud components and the HD\,130079 starlight separately (top and bottom panels of Figure\,\ref{fig:SED}, respectively). For the microwave segment of the cloud SED, dominated by cold dust, we take aperture fluxes from the Second Planck Catalogue of Compact Sources \citep[PCCS2E;][]{PCCS2E16} at 143, 217, 353, 545, and 857\,GHz. In the infrared range, dominated by the radiative output of the warm dust in the cloud's regions adjacent to the star, we use fluxes from the IRAS Point Source Catalog v.2.1 at 12, 25, 60, and 100\,$\mu$m, and the WISE fluxes at 3.4, 4.6, 12, and 22\,$\mu$m, all corresponding to the infrared source IRAS\,14451--6502 associated with HD\,130079 (see Section\,\ref{sec:IRAS}). Indeed as seen in Figure\,\ref{fig:low-res}, the IRAS and (to a lesser extent) the Planck images display a shifted maximum peak away from the center of DC\,314.8--5.1. The different apertures and resolutions of the Planck ($5^{\prime}$), IRAS ($1^{\prime}$), and WISE ($6-12^{\prime\prime}$) instruments could be important though. These differences are particularly relevant in the case of the IRAS vs. WISE comparison, and could explain the lower WISE fluxes when compared to the IRAS photometry within the overlapping wavelength range of 12--25\,$\mu$m.

We have calculated several model curves for the cold thermal component using a modified blackbody emission $B_{\nu}\!(T) \times (\nu/\nu_0)^{\beta}$. Our findings indicate that the best-matching model corresponds to a temperature of $T=14$\,K, consistent with the PGCC model fit, and a spectral index of $\beta = 1.5$ (see the dark red solid curve in Figure\,\ref{fig:SED}; cf. Section\,\ref{sec:Planck}). The far/mid-infrared (roughly $7-70$\,$\mu$m) emission of the system is a complex superposition of the continua from multi-temperature dust, molecular lines, and PAH features, all generally decreasing in intensity with distance away from the photoionizing star DC\,314.8--5.1 \citep[see][]{Kosmaczewski22}. As a basic representation of the entire spectral component, we adopt the simplest model, which consists of a single modified blackbody. This time, the model has a temperature of 160\,K, a spectral index of $\beta=2.0$, and a normalization adjusted to match the 12--25\,$\mu$m IRAS fluxes (see the dark red dashed curve in Figure\,\ref{fig:SED}).

While the cold component temperature is precisely constrained by the multiwavelength (143--857\,GHz) Planck data in conjunction with the IRAS 100\,$\mu$m photometry, the warm component's temperature lacks such precision. As previously emphasized, using a single modified blackbody to approximate the hot dust emission in the system is a basic, zero-order approximation. It is worth noting that the IRAS 60\,$\mu$m flux, which surpasses both the cold (14\,K) and warm (160\,K) blackbody emission components, indicates the presence of gas with intermediate temperatures in the system. As such, this model is meant to be primarily illustrative.

For the SED representing the HD\,130079 starlight, the near-infrared (filters $JHKL$) and optical ($UBV$) fluxes follow directly from the compilation by \citet[][see Table\,1 therein]{Whittet07}, with the addition of the $G$ band flux from the EDR3 \citep{Gaia21}, and the UV 2250\AA\, flux measured from the newly obtained Swift UVOT observations (see Section\,\ref{sec:Swift}). The photospheric emission of HD\,130079, is modeled here assuming a simple optically-thick blackbody spectrum with the temperature $T_{\star} = 10,500$\,K, such that the bolometric stellar luminosity is $L_{\star} = 4\pi R_{\star}^2 \, \sigma_{\rm SB} T_{\star}^4 \simeq 3 \times 10^{35}$\,erg\,s$^{-1}$, for the stellar radius $R_{\star} = 2.7 \times R_{\odot}$, and the distance of 432\,pc. This intrinsic emission (denoted in Figure\,\ref{fig:SED} by the dark blue dot-dashed curve) is next reduced by interstellar reddening using the \citet{Cardelli89} empirical extinction law with the coefficients as given in equations 2--5 of \citeauthor{Cardelli89}, and values for $E_{B-V}$ ($ = 0.395$) and $R_V$ ($= 4.5$, in excess over the averaged ISM value of $3.1$) adopted from \citet{Whittet07}. The reddened starlight (given by the solid dark blue curve in Figure\,\ref{fig:SED}), matches the near-infrared--to--UV fluxes of the star including the 3.4 and 4.6\,$\mu$m WISE fluxes, and the Swift UVOT 2250\AA\, flux, even though no stellar photospheric reddening was included in this simple model.

The mass of the Planck source PGCC\,G314.77--5.14, has been estimated in \citet{Planck16} as
\begin{equation}
M = \frac{D^2 \, F_{\nu}}{\kappa_{\nu} \, B_{\nu}\!(T)} \simeq 10\pm 14\,M_{\odot},
\label{eq:massPlanck}
\end{equation}
based on the measured 857\,GHz flux density $F_{\nu}$ integrated over the solid angle $\Omega = \pi \theta^2/4$ (where $\theta$ is the geometric mean of the major and minor FWHM), which is effectively half the provided PGCC flux, with the dust opacity value $\kappa_{\nu} =0.1\, (\nu/1\,{\rm THz})^2$\,cm$^{2}$\, g$^{-1}$ adopted from \citet{Beckwith90}. Meanwhile, \citet{Whittet07} estimated the mass of the \emph{core} of the globule to be $\gtrsim 50\,M_{\odot}$ when updated for the 432\,pc distance. However, this discrepancy might not be significant, keeping in mind that the Planck estimate provides upper $2\sigma$ (95\%) and $3\sigma$ (99\%) confidence limits of $68\, M_{\odot}$ and $115\,M_{\odot}$, respectively. These limits arise solely from uncertainties in flux and distance estimates, and do not account for the uncertainty in the dust opacity function, $\kappa_{\nu}$ (see in this context the discussion in \citealt{Beckwith90}, specifically Section IIIe, and also \citealt{DAlessio01}).

The mass of the cloud --- as an isolated dark cloud at high Galactic latitudes --- can also be estimated from the excess absorption seen in X-rays toward the cloud \citep[see in this context][]{Sofue16} and from the high-energy $\gamma$-ray data as measured by the Fermi's Large Area Telescope \citep[LAT; see][and references therein]{Mizuno22}. In the former case, a much deeper X-ray observations would be needed to estimate the absorbing hydrogen column density across the cloud. Concerning the latter, we note that in the Fermi High-Latitude Extended Sources Catalog (FHES) by \citet{Ackermann18}, the integrated 1\,GeV--1\,TeV fluxes of resolved high confidence sources in the LAT data extend down to a few/several $\times 10^{-10}$\,cm$^{-2}$\,s$^{-1}$. Further, those which appear point-like lie about one magnitude lower, with a median of $2.5 \times 10^{-10}$\,cm$^{-2}$\,s$^{-1}$. An estimate for the flux expected from DC\,314.8--5.1 due to the
interactions with high-energy CRs (assuming no CR overdensity with respect to the CR background), is
\begin{eqnarray}
F(>E_{\gamma}) & \simeq & 2 \times 10^{-13} \, \frac{M/10^5\,M_{\odot}}{(D/{\rm kpc})^2}\,\left(\frac{E_{\gamma}}{1\,{\rm TeV}}\right)^{-1.7} \nonumber \\
& \sim & 2 \times 10^{-10}\,{\rm cm^{-2}\,s^{-1}}
\label{eq:gamma}
\end{eqnarray}
\citep[see][]{Gabici13}. In the above, we use $M=160\,M_{\odot}$ corresponding to the \emph{total} mass of the cloud \citep[updated distance $D=432$\,pc]{Whittet07}, and $E_{\gamma} = 1$\,GeV. This level of emission may be detected in dedicated Fermi-LAT studies, leading to a robust estimate of the mass in this pre-stellar, condensed dark cloud.

\section{Conclusions}

 In this paper we have discussed the multi-wavelength properties of the dark globule, DC\,314.8--5.1, through dedicated observations with the Spitzer Space Telescope and the Swift-XRT and UVOT instruments, supplemented by the archival Planck, IRAS, WISE, 2MASS, and Gaia data. This investigation of the characteristics of the system, over a wide range of the electromagnetic spectrum, has led to the following conclusions:
\begin{enumerate}
    \item We have further supported that DC\,314.8--5.1 is a pre-stellar core, with no conclusive Class\,I-III YSO candidates present within the extent of the system down to luminosities as low as $\sim 10^{-4} L_{\odot}$, translating to a stellar core mass down to $0.01 M_{\odot}$ \citep[see][]{Dunham12}. We do, however, maintain on possible candidate Class\,III YSO object (``C2,'' see Table\,\ref{table:S}), albeit unlikely due to the lack of heating seen in that region of DC\,314.8--5.1. Furthermore, we exclude any younger Class 0 YSO candidates down to luminosities of $\sim 0.18 \, L_{\odot}$ translating to a core mass of $\sim 0.1 \, M_{\odot}$ \citep[see in this context][]{Barsony97}.
    
    \item With the Swift-XRT observations, we probed for any potential PMS population down to a luminosity level of  $\lesssim 10^{31}$\,erg\,s$^{-1}$. This level would have detected a typical PMS of mass\,$\ge 2 M_\odot$, while being capable of only detecting the brightest (flaring) low-mass ($< 2 M_\odot$) PMSs. Deeper observations would be needed to reject the presence of lower mass objects. Furthermore, CO observations of this system could also test for the presence of the youngest, Class 0, protostars \citep[][and references therein]{Kirk05}.
    
    \item We investigated the SED of the DC\,314.8--5.1 system as well as the nearby illuminator HD\,130079. Our analysis confirmed the presence of warm dust, with temperatures $\gtrsim 100$\,K, in addition to the dominant 14\,K dust component. This warm component manifests itself in the IRAS photometry, particularly within the 12-25\,$\mu$m range.
    
    \item We comment on the variation in mass estimates of DC\,314.8--5.1, ranging from $\simeq 12\,M_{\odot}$ based on the Planck photometry, up to $\gtrsim 50\,M_{\odot}$, following from the visual extinction characteristics, for the core of the globule. We point out that the discrepancy may be due to errors in the flux measurement, variations in the methodology, and the opacity model uncertainties for this particular system. We also note in this context, that the cloud should be detectable in high-energy $\gamma$-rays with Fermi-LAT, given the estimate for the total mass of the globule $\sim 160\,M_{\odot}$.
\end{enumerate}
Hence, DC\,314.8--5.1 remains a pre-stellar cloud core. This makes it an ideal candidate for deeper observations, particularly in high-energy X-ray and $\gamma$-ray.

%\clearpage

\begin{acknowledgments}
EK and \L .S. were supported by Polish NSC grant 2016/22/E/ST9/00061. This research was completed while EK held an NRC Research Associateship award at the Naval Research Laboratory. WRMR thanks the financial support from the Leiden Observatory. AK acknowledges support from the First TEAM grant of the Foundation for Polish Science No. POIR.04.04.00-00-5D21/18-00 and the Polish National Agency for Academic Exchange grant No. BPN/BEK/2021/1/00319/DEC/1. C.C.C. was supported by NASA DPR S-15633-Y. WRMR thanks the financial support from the European Research Council
(ERC) under the European Union’s Horizon 2020 research and innovation
programme (grant agreement No. 101019751 MOLDISK). The authors thank the anonymous referee for their comments and suggestions which helped to improve the manuscript.

This work is based on observations made with the Spitzer Space Telescope, obtained from the NASA/ IPAC Infrared Science Archive, both of which are operated by the Jet Propulsion Laboratory, California Institute of Technology under a contract with the National Aeronautics and Space Administration. 

This work has made use of data from the European Space Agency (ESA) mission Gaia (\url{https://www.cosmos.esa.int/gaia}), processed by the Gaia Data Processing and Analysis Consortium (DPAC, \url{https://www.cosmos.esa.int/web/gaia/dpac/consortium}). Funding for the DPAC has been provided by national institutions, in particular the institutions participating in the Gaia Multilateral Agreement. We are grateful to Timo Prusti for advice on Gaia data.

This research used data products from the Two Micron All Sky Survey, a joint project of the University of Massachusetts and the Infrared Processing and Analysis Center, funded by the National Aeronautics and Space Administration and the National Science Foundation. The Digitized Sky Survey was produced at the Space Telescope Science Institute under U.S. Government grant NAG W-2166. The images of these surveys are based on photographic data obtained using the Oschin Schmidt Telescope on Palomar Mountain and the UK Schmidt Telescope. The plates were processed into the present compressed digital form with the permission of these institutions.

\end{acknowledgments}

\appendix

\section{Mid-Infrared Images of DC 314.8--5.1}
\label{A:MIR-images}

\begin{figure*}[th!]
\centering
\includegraphics[width=\textwidth]{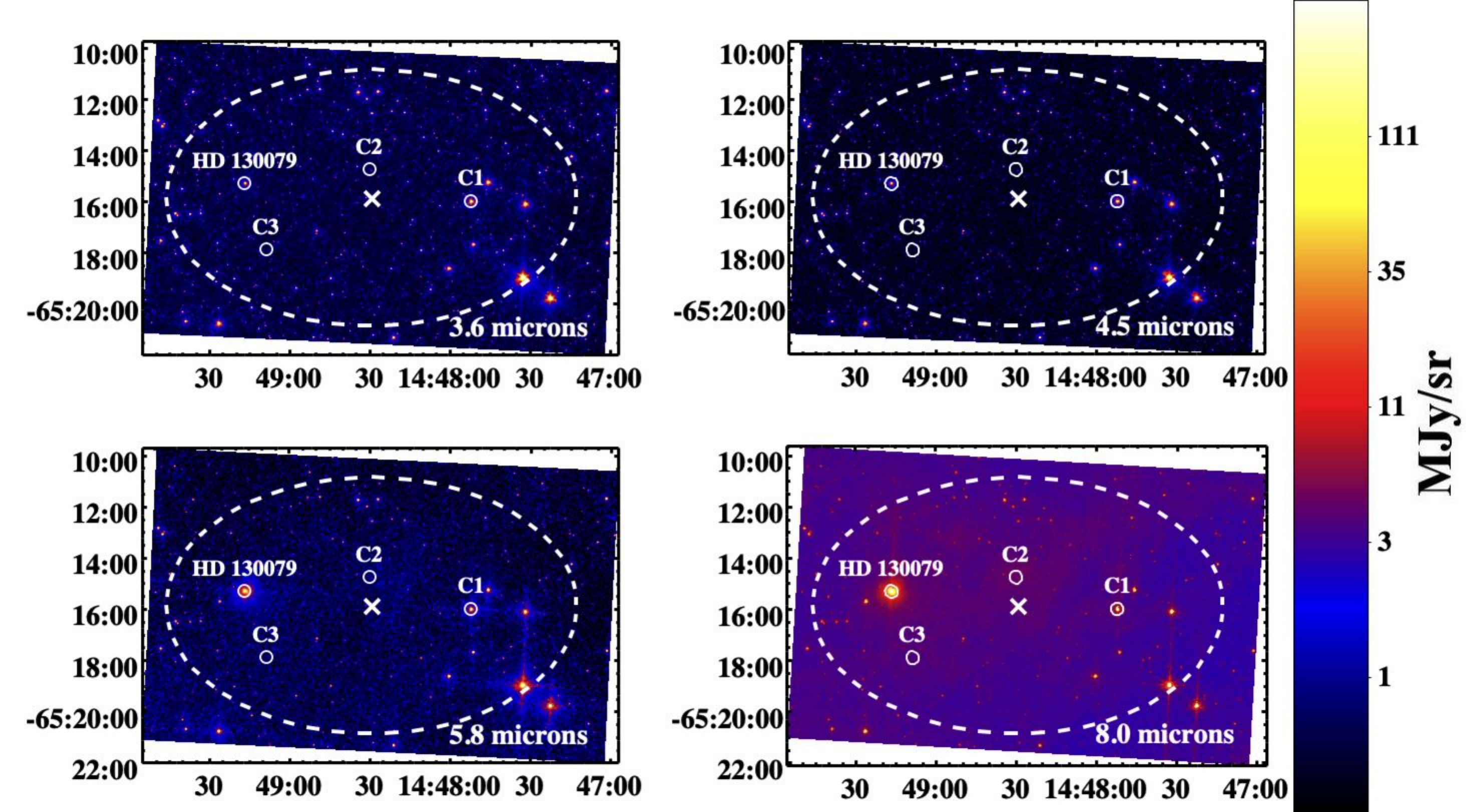}
\caption{Spitzer IRAC band maps of the DC\,314.8--5.1 region. (top left) Channel 1 at 3.6\,$\mu$m, (top right) channel 2 at 4.5\,$\mu$m, (bottom left) channel 3 at 5.8\,$\mu$m, and (bottom right) channel 4 at 8.0\,$\mu$m. White dashed ellipse marks the optical extent of DC\,314.8--5.1. Color bar is set with a minimum of zero and a maximum value of 200 MJy/sr on a log scale. The HD\,130079 field star is marked in the left of each image, ``C1'' marks the YSO candidate identified by \citet{Whittet07}, and ``C2'' marks the YSO candidate identified in this work.} 
\label{fig:IRAC}
\end{figure*}

In Figure\,\ref{fig:IRAC}, we present the dedicated Spitzer IRAC images of the DC\,314.8--5.1 region at  3.6, 4.5, 5.8, and 8.0\,$\mu$m with a spatial resolution of $\sim 2\arcsec$ \citep{Fazio04}. As is visible in the figure, the superior spatial resolution of the Spitzer instrument allows for the detection of individual point sources within DC\,314.8--5.1. Additionally, the diffuse emission of the cloud and some faint irregular structure can be seen in the bottom right panel of Figure\,\ref{fig:IRAC}. 

\section{ROSAT and Swift XRT Comparison}
\label{A:ROSAT}

\begin{figure*}[h!]
\centering
\includegraphics[width=0.45\textwidth]{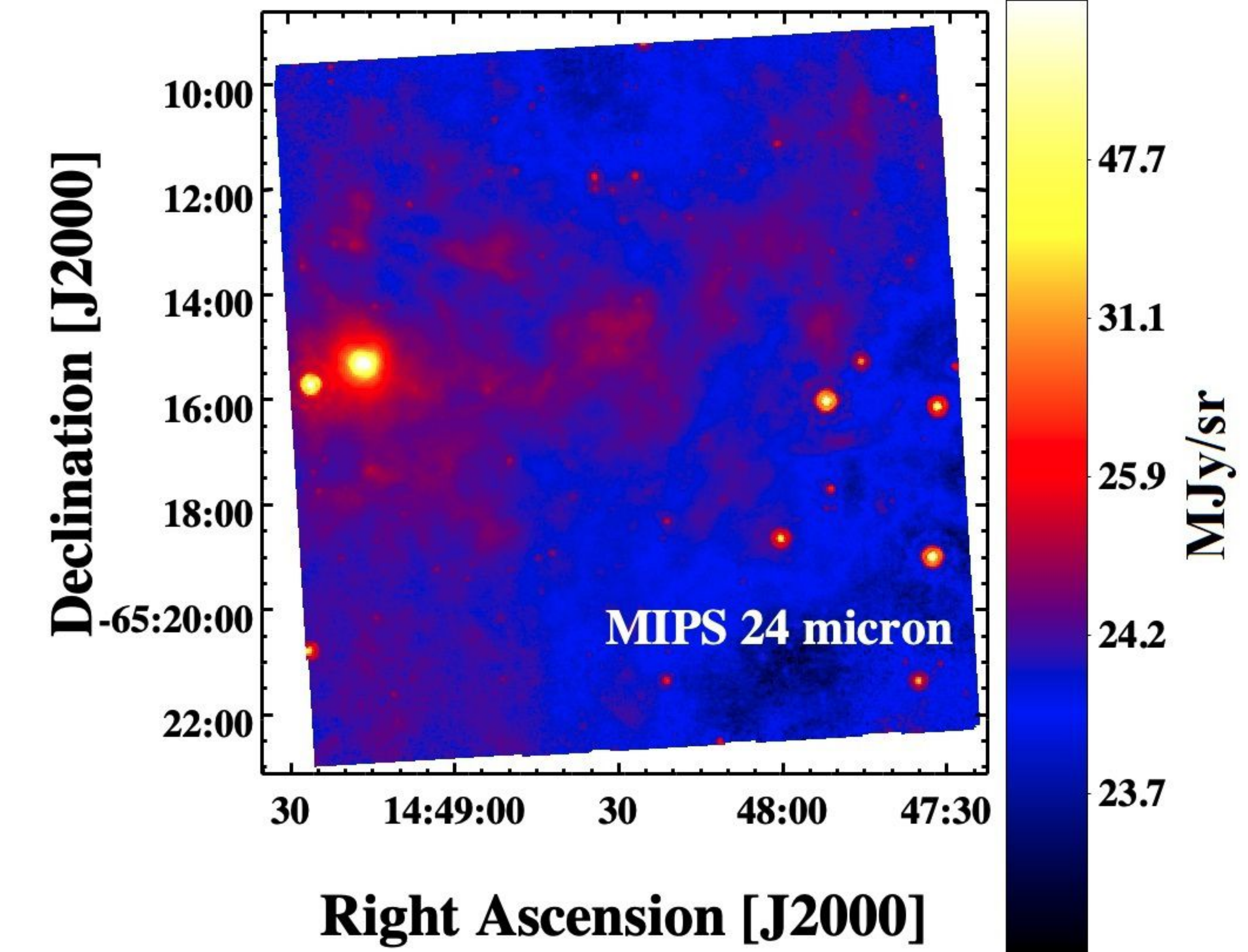}
\includegraphics[width=0.45\textwidth]{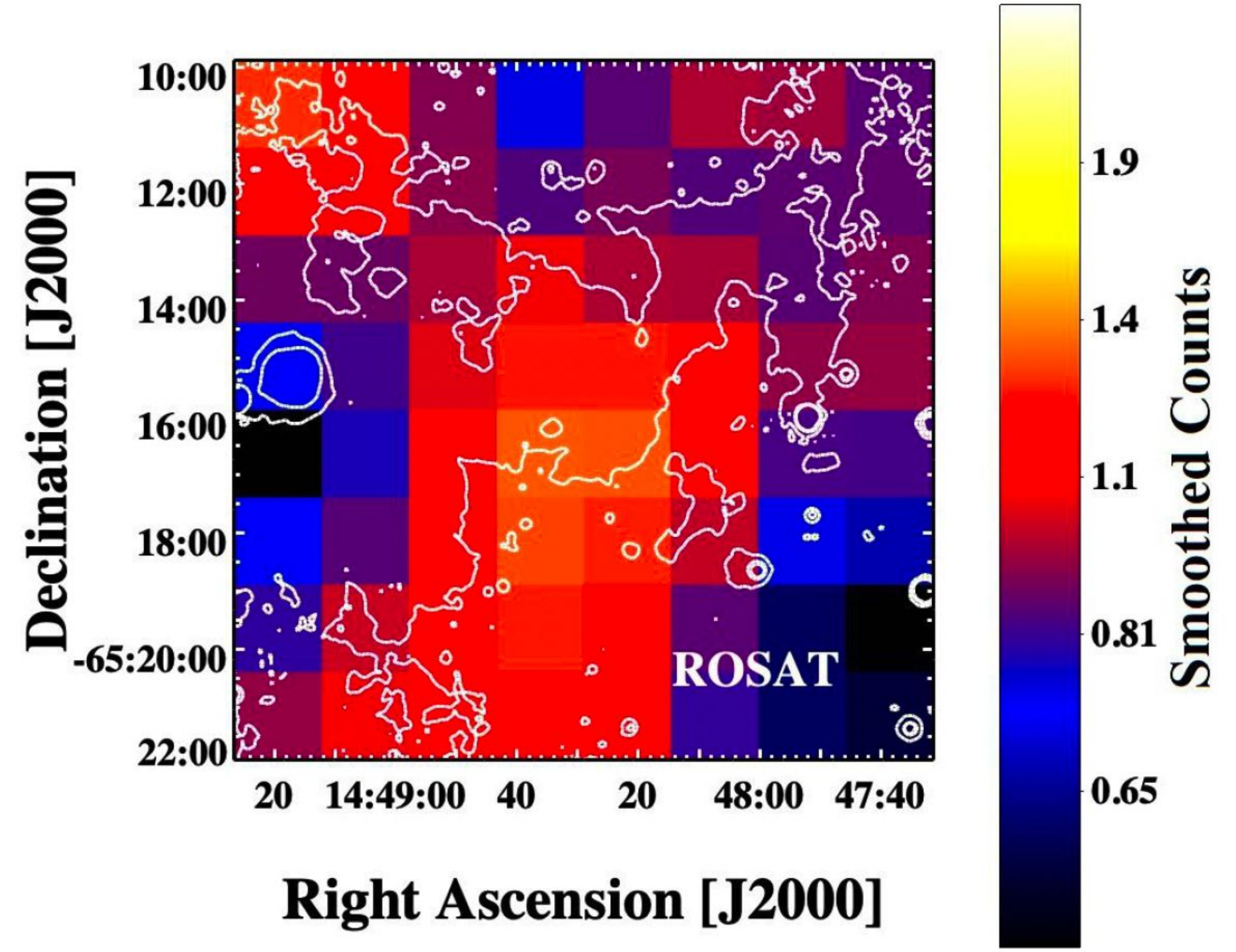}\\
\includegraphics[width=0.475\textwidth]{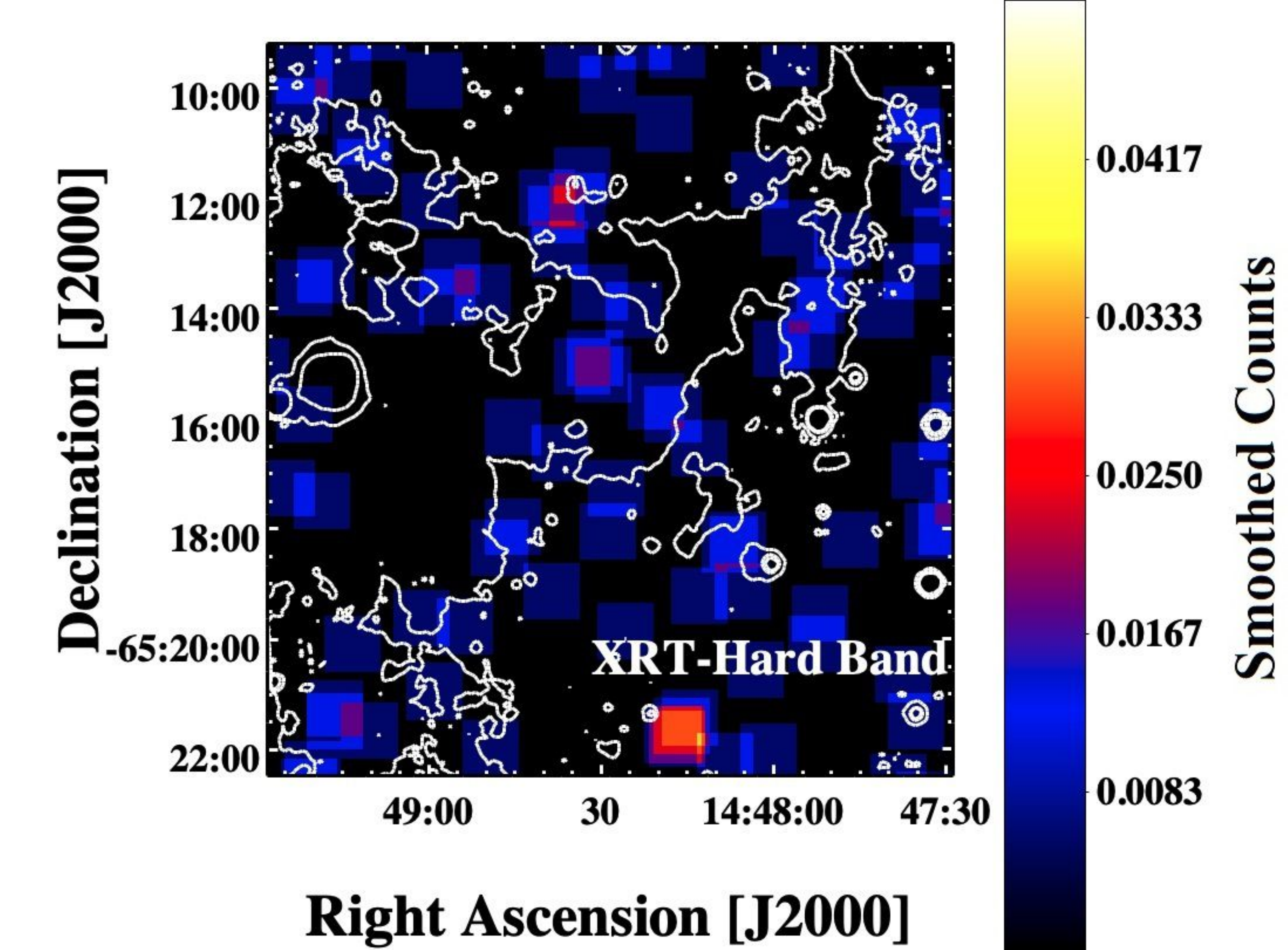}
\includegraphics[width=0.475\textwidth]{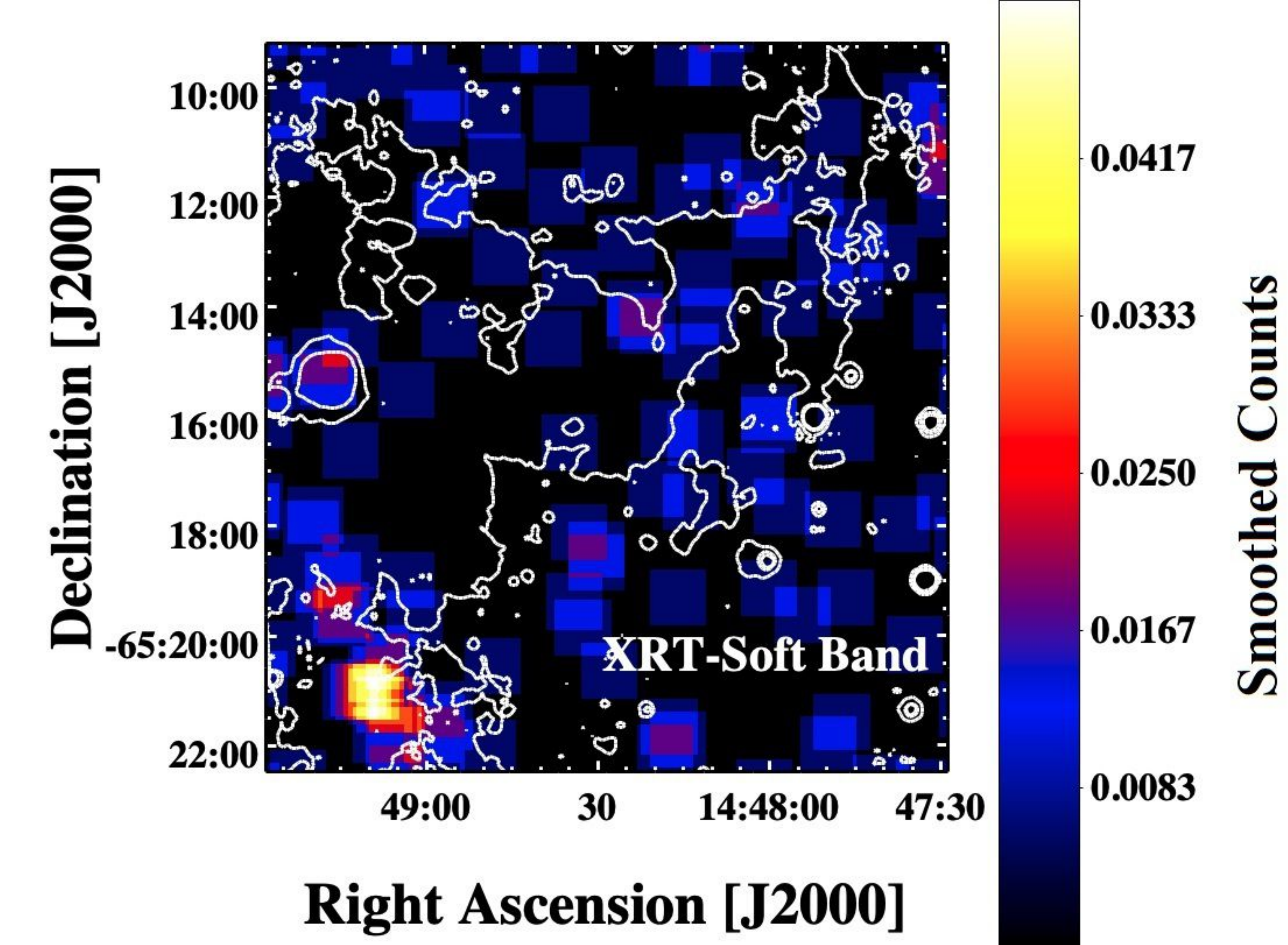}
 \caption{(top left) Spitzer MIPS 24\,$\mu$m image of the DC\,314.8--5.1 region. Color bar spans from a minimum of 23.5 to 100\,MJy/sr on a log scale. White contours in the following panels represent MIPS emission, with levels set at 24.1, 25.05, and 26\,MJy/sr. (top right) ROSAT full-band $0.1-2.4$\,keV image of the same region, with Spitzer MIPS contours superimposed. Color bar shows a range 0.6--2.5 smoothed counts on a sqrt scale with a Gaussian smoothing, see Section\,\ref{A:ROSAT}. (bottom) Swift XRT hard-band $2-10$\,keV and soft-band $0.3-2.0$\,keV images of the same region (left and right, respectively), smoothed with a boxcar kernal, see Section\,\ref{A:ROSAT}. Spitzer MIPS contours are superimposed in white for reference. Color bar shows a range of 0--0.05 smoothed counts on a linear scale. }
\label{fig:ROSAT-XRT}
\end{figure*}

In Figure\,\ref{fig:ROSAT-XRT}, we compare the ROSAT full-band $0.1-2.4$\,keV image of the DC\,314.8--5.1 region with the corresponding images from the Swift XRT soft band $0.3-2.0$\,keV as well as hard band $2-10$\,keV, including for reference the MIPS 24\,$\mu$m image of the region. We note that, assuming the emission model specified in Section\,\ref{sec:PMS}, the PSPC $0.1-2.4$\,keV count rate for ROSAT J144833.7--651738 of $0.0175\pm 0.0082$\,cts\,s$^{-1}$, would correspond to the Swift XRT PC  $0.3-10$\,keV count rate of $(19.7\pm 9.2) \times 10^{-3}$\,cts\,s$^{-1}$ for $N_{\rm H,\,int} \simeq 10^{22}$\,cm$^{-2}$, and $(13.1\pm 6.2) \times 10^{-3}$\,cts\,s$^{-1}$ for $N_{\rm H,\,int} \simeq 10^{21}$\,cm$^{-2}$, while the estimated XRT $0.3-2.0$\,keV count rate for the source ``S'' is $4.1^{+1.8}_{-1.4} \times 10^{-3}$\,cts\,s$^{-1}$, and the $3\sigma$ upper limit at the nominal position of the center of the cloud/ROSAT source, calculated with The Living Swift XRT Point Source Catalogue online tool\footnote{\url{https://www.swift.ac.uk/LSXPS/}}, is $\simeq 3\times 10^{-3}$\,cts\,s$^{-1}$. The Swift-XRT images have been smoothed for visualization purposes to more closely resemble the resolution of the ROSAT observation using a boxcar with a width\,$=2r+1$\,pixels and radius (r)\,$=3$\,pixels \citep{Joye03}.

\section{IRAC--2MASS Color Cuts}
\label{A:IRAC-cuts}

\begin{figure*}[th!]
    \centering
    \includegraphics[width=0.49\textwidth]{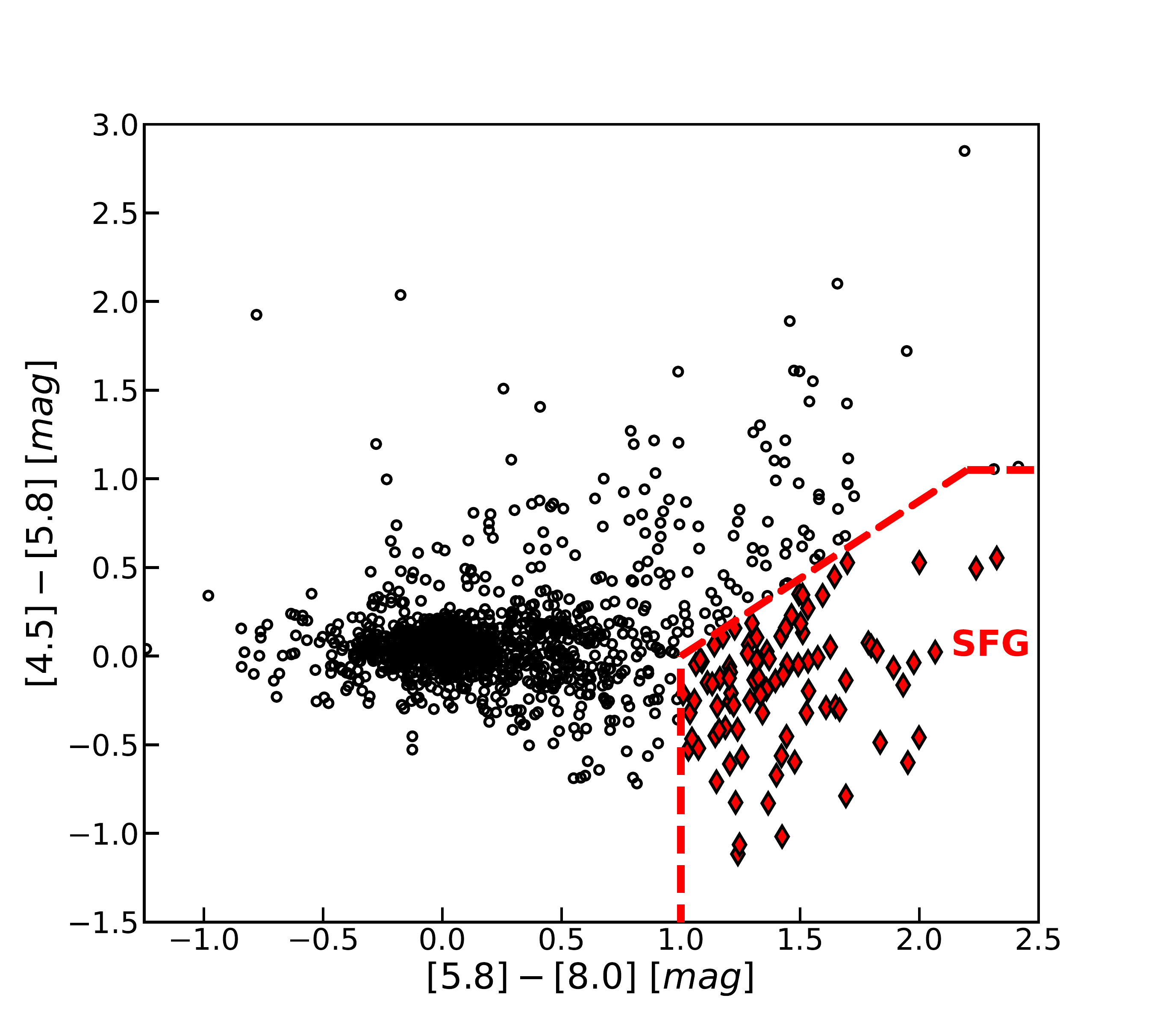}
    \includegraphics[width=0.49\textwidth]{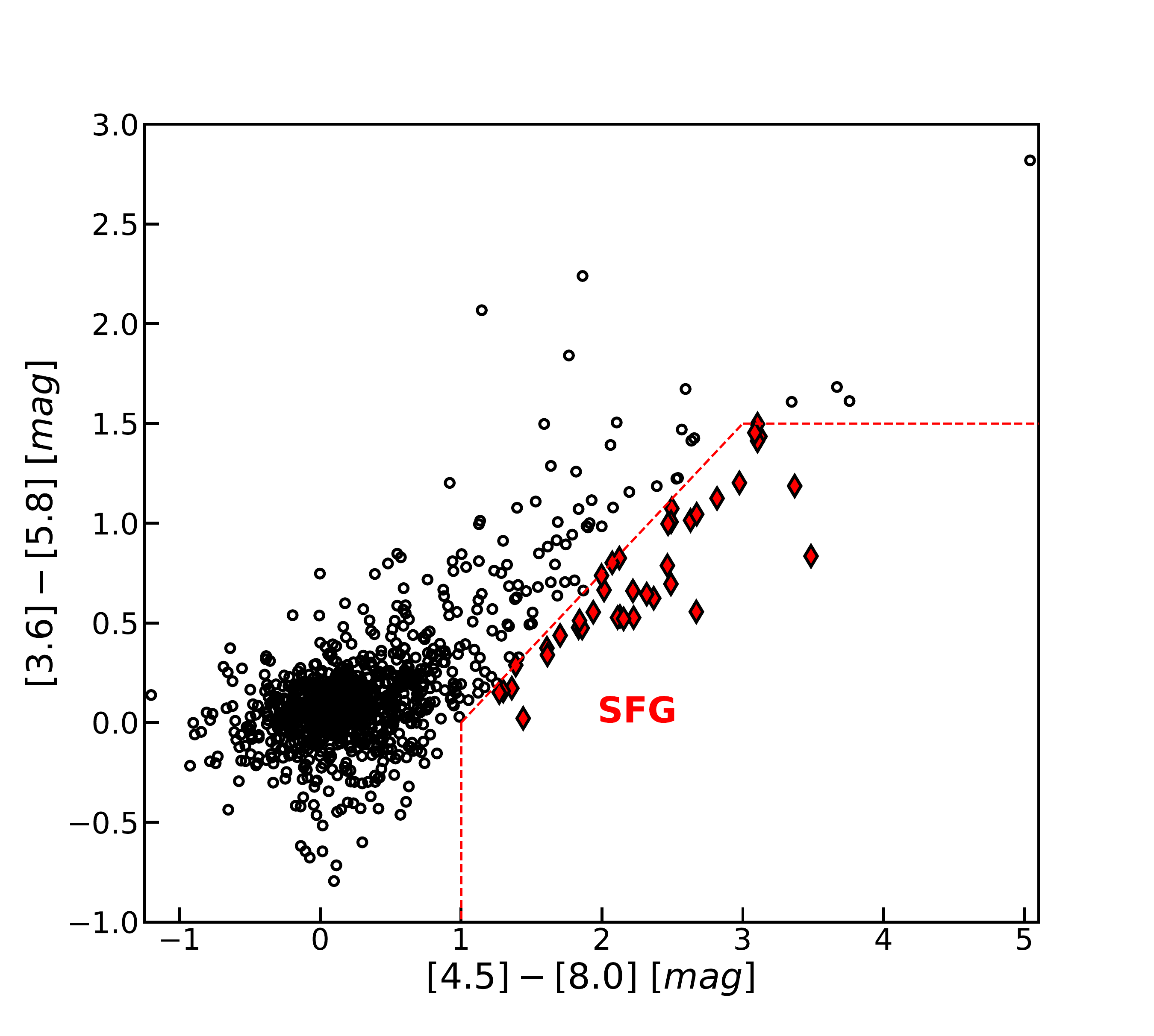}\\ \vspace{-0.5cm}
    \includegraphics[width=0.49\textwidth]{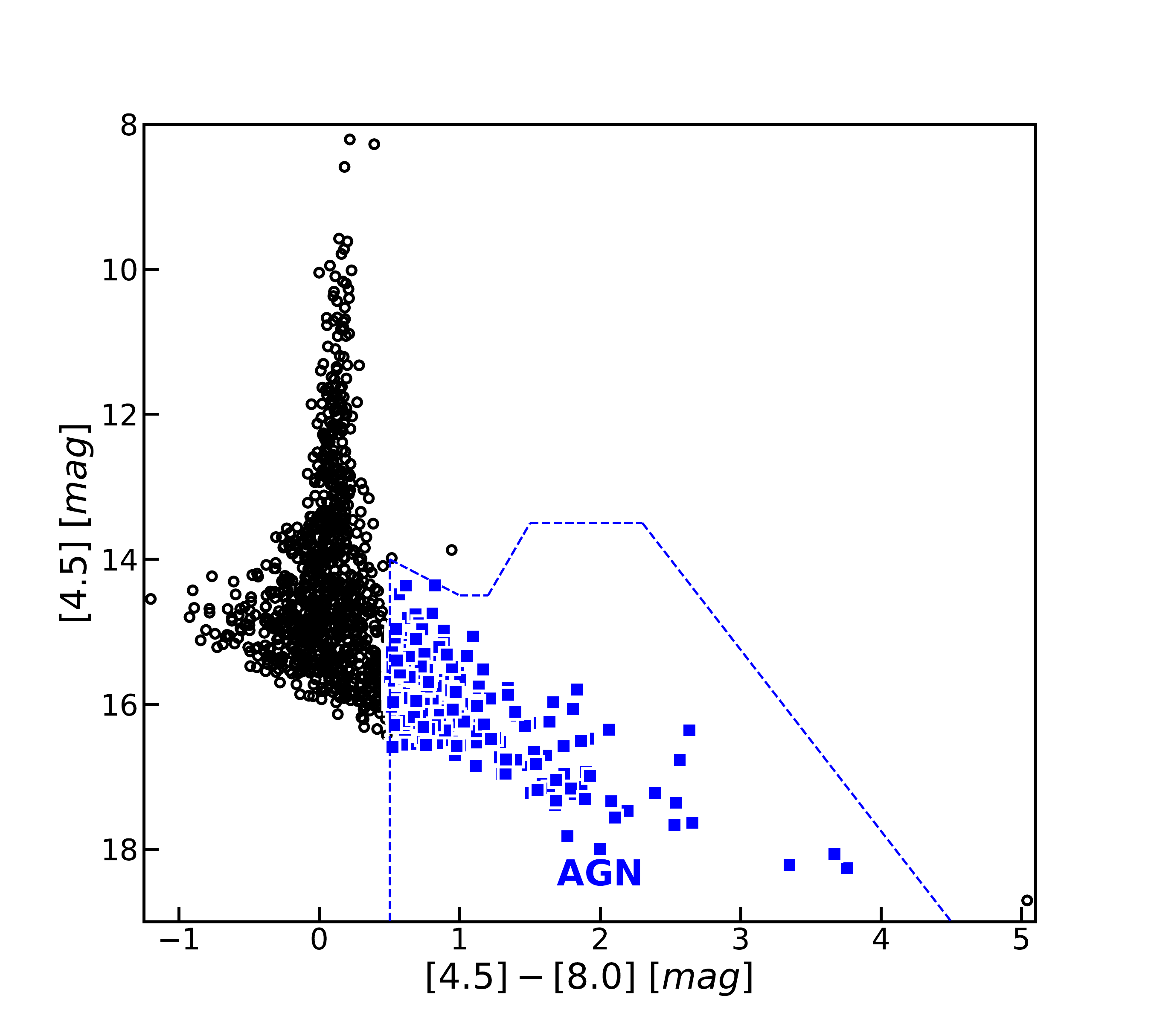}
    \includegraphics[width=0.49\textwidth]{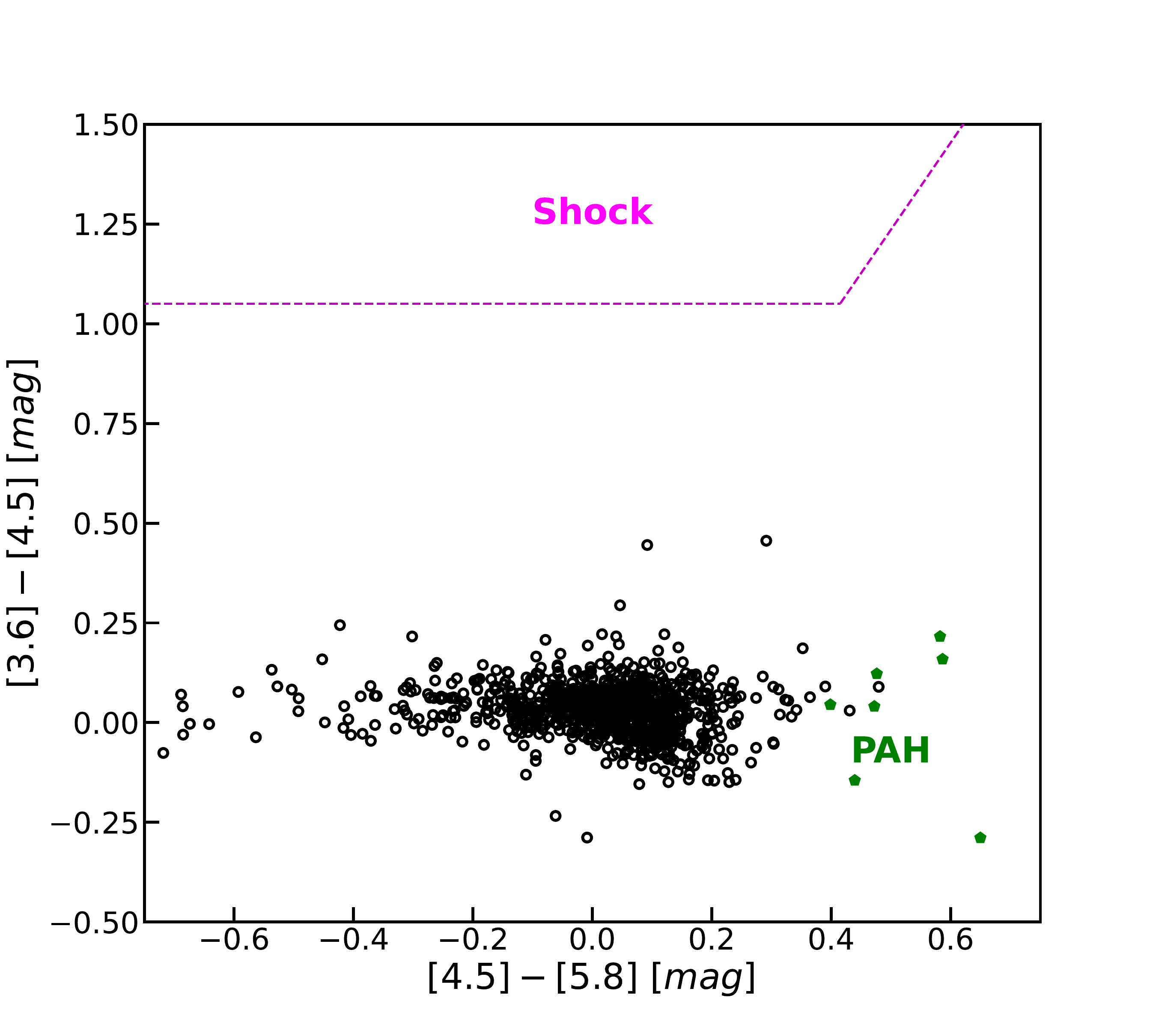}\\ \vspace{-0.5cm}
    \caption{IRAC color-color diagram with selection cuts from \citet{Gutermuth09}. (top left) Total 1,319 sources from the SEIP query of 5$^\prime$ radius from the optical center of DC\,314.8--5.1. Overlaid with the first selection cut, red dashed lines showing the removal of star forming galaxies (SFG). (top right)  Second SFG selection cut, removed sources are marked as red diamonds. (bottom left) AGN selection cut with AGN like sources marked in blue. (bottom right) Selection cut in order to remove in order to remove Galactic-scale unresolved shocks and unresolved PAH emission sources, PAH emission sources are marked with green pentagons, no unresolved shock sources were selected.} 
    \label{fig:IRAC-cuts}
 \end{figure*}

\begin{figure*}[th!]
    \centering
    \includegraphics[width=0.7\textwidth]{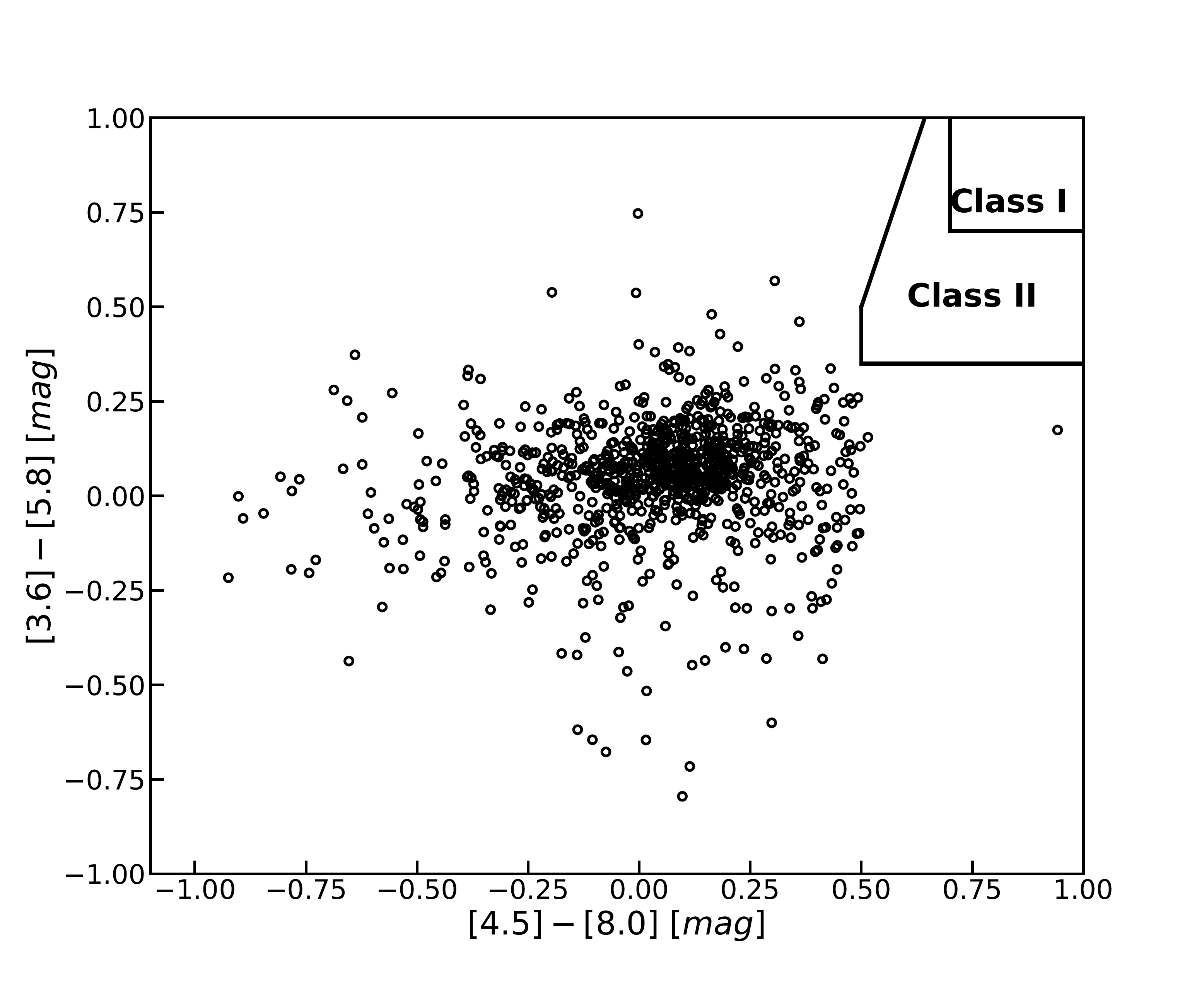}
    \caption{IRAC color-color diagram with YSO selection criteria from \citet{Gutermuth09}. Black marked regions display the final selection cut identifying YSO Class I and II sources. Sources outside the delineated regions are undefined in this selection. } 
    \label{fig:IRAC-YSO}
 \end{figure*}

We utilize the criteria presented in \citet{Gutermuth09} and \citet{Winston19} to select YSO candidates in DC\,314.8--5.1 based on Spitzer IRAC mapping data, as outlined in Section\,\ref{sec:IRAC-YSO}. The top left panel of Figure\,\ref{fig:IRAC-cuts} displays the full extent of the Spitzer sample in the region detecting 1,319 sources within $5^{\prime}$ of the center of DC\,314.8--5.1 as well as the first star-forming galaxy selection cut. Figure\,\ref{fig:IRAC-cuts} further displays the selections done on the sample to remove contaminating sources beginning with removal of star-forming galaxies (top panels), followed by AGN and finally unresolved PAH and shock emission (bottom left and right panels, respectively). We further show the final selection, identifying Class\,I and Class\,II YSOs in Figure\,\ref{fig:IRAC-YSO}. Sources that fall outside the YSO selection regions (923 sources) are unidentified in the selection, and assumed to fall under the Class\,III or Field Star category. Given these remaining sources are heavily contaminated with AGB stars in similar samples, see \citet{Dunham15}, we further cross-correlated with Gaia measured distances and further discussed in Section\,\ref{sec:Gaia} and \ref{sec:discussion}. 
\end{document}